\documentclass[10pt]{iopart}
\usepackage{graphicx,amsfonts,amssymb,color}
\expandafter\let\csname equation*\endcsname\relax
\expandafter\let\csname endequation*\endcsname\relax
\usepackage{amsmath}
\usepackage{dsfont}
\usepackage[normalem]{ulem}

\usepackage[english]{babel}
\usepackage{fancybox}
\usepackage{verbatim}
\usepackage{rotating}
\usepackage{sidecap}
\usepackage[caption=false]{subfig}
\usepackage{hyperref}

\newcommand{\p}[2]{\frac{\partial #1}{\partial #2}\,}
\newcommand{\intOm}[1]{\int\limits_\Omega #1\rmd \V{r}\,}

\newcommand{\intT}[1]{\int\limits_{0}^{T} #1\rmd t\,}
\newcommand{\intQ}[1]{\intT{\hspace*{-5pt}\intOm{#1}}}

\newcommand{\lao}{L^1(\Omega)}

\newcommand{\Z}{\mathbb{Z}}
\newcommand{\R}{\mathbb{R}}
\newcommand{\V}[1]{\mathbf{#1}}
\newcommand{\norm}[1]{\lVert#1\rVert}
\newcommand{\bracket}[1]{\left(#1\right)}

\newcommand{\Vthree}[3]{\begin{pmatrix} #1 \\ #2  \\ #3 \end{pmatrix}}

\newcommand{\grb}[1]{\mbox{\boldmath $#1$}}

\newcommand{\eq}[1]{\eqref{#1}}
\newcommand{\eqs}[1]{\eqref{#1}}
\newcommand{\figref}[1]{figure~\ref{#1}}
\newcommand{\secref}[1]{section~\ref{#1}}
\newcommand{\tabref}[1]{table~\ref{#1}}

\newcommand{\new}[1]{{\color{red} [\textbf{NEW}: #1]}}

\begin{document}
\title[Control of traveling localized spots]{Control of traveling localized spots}
\author{S. Martens$^1$, C. Ryll$^2$, J. L\"ober$^{1,3}$, F. Tr{\"o}ltzsch$^2$, and H. Engel$^1$} 
\address{$^1$ Institut f\"ur Theoretische Physik, Hardenbergstra\ss e 36, EW 7-1, Technische Universit\"at Berlin, 10623 Berlin, Germany}
\address{$^2$ Technische Universit\"at Berlin, Institut f\"ur Mathematik, Str. d. 17. Juni 136, MA 4-5, 10623 Berlin, Germany}
\address{$^3$ Max-Planck-Institut f\"ur Physik komplexer Systeme, N\"othnitzer Stra\ss e 38, 01187 Dresden, Germany}
\ead{\mailto{steffen.martens@tu-berlin.de}}

\begin{abstract}
Traveling localized spots represent an important class of self-organized two-dimensional patterns in reaction-diffusion systems. We study open-loop control intended to guide a stable spot along a desired trajectory with desired velocity. Simultaneously, the spot's concentration profile does not change under control. For a given protocol of motion, we first express the control signal analytically in terms of the Goldstone modes and the propagation velocity of the uncontrolled spot. Thus, detailed information about the underlying nonlinear reaction kinetics is unnecessary. Then, we confirm the optimality of this solution by demonstrating numerically its equivalence to the solution of a regularized, optimal control problem. To solve the latter, the analytical expressions for the control are excellent initial guesses speeding-up substantially the otherwise time-consuming calculations.
\end{abstract}

\pacs{02.30.Jr, 02.30.Yy, 82.40.Bj, 82.40.Ck}

\noindent{\it Keywords\/}: open-loop control, optimal control, reaction-diffusion system, spots, coherent structures

\submitto{\NJP}
\maketitle

\section{Introduction}

Localized spots, sometimes referred to as auto-solitons \cite{kerner2013autosolitons}, dissipative solitons \cite{purwins_dissipative_2005}, or bumps \cite{Laing2002}, are a subclass of traveling waves that spontaneously evolve in two-dimensional (2D) dissipative nonlinear systems far from thermodynamic equilibrium. In a co-moving reference frame, spots are stationary solutions to coupled nonlinear partial differential equations (PDE), such as reaction-diffusion (RD) or neural field equations, for example. The characteristic length and time scales of the spots, i.e., their wave profile, propagation velocity, etc., are selected by the experimental conditions or the parameters of the model.

\indent Experimentally, localized spots have been observed as current filaments in gas-discharge \cite{Purwins2010}, as bright intensity spots in nonlinear optics and laser physics \cite{arecchi_pattern_1999}, as moving localized regions of increased concentration in chemical reactions \cite{vanag_localized_2007}, or coverage of adsorbed species in heterogeneous catalysis \cite{wolff2001spatiotemporal}. Further examples include  temperature spots in fixed-bed catalytic reactors \cite{Sheintuch2008}, actin conformation in dictyostelium discoideum \cite{le_goff_pattern_2016}, neural activity in head-direction cells \cite{taube_persistent_2003}, vegetation patterns \cite{gilad_ecosystem_2004}, and many others.

Although control of self-organized patterns attracted considerable attention over the last decades, compare \cite{Mikhailov2006,vanag_design_2008} and references therein, it is still a challenging problem in applied nonlinear science. Often, one distinguishes between open-loop, closed-loop, and optimal control.\\
Open-loop control is independent of the instantaneous state of the system. As a consequence, it is inherently susceptible to perturbations in the initial conditions as well as to parameter uncertainty. Thus, detailed knowledge of the system's dynamics and in-depth stability analysis are pre-conditions for reliable open-loop schemes. Typical examples of open-loop control are space-time dependent external forcing \cite{Zykov2003,Chen2009} or control by imposed geometric constraints \cite{Haas1995,ziepke_wave_2016}.\\
On the other hand, in closed-loop or feedback control, the controlled state is permanently monitored to adjust the control signal accordingly \cite{pierre_controlling_1996,luthje_control_2001,Kim2001}. Particularly, time-delayed feedback can induce pattern forming instabilities in addition to the pattern to be controlled \cite{gurevich_instabilities_2013}.\\
Optimal control reformulates control problems in terms of the minimization of a cost functional \cite{bryson1975applied,hllst02}. The cost functional measures the distance in function space between a desired target state and the actual controlled state of the system. If a control signal is the unique solution to an optimal control problem, then no other control, be it open- or closed-loop, will be able to enforce a controlled state closer to the target state. Conditions for sufficiency and uniqueness of optimal control are discussed extensively in the mathematical literature \cite{hlt98,bjt10,casas_ryll_troeltzsch2014}. Optimal control of self-organized patterns requires complete knowledge of the PDEs governing the system's evolution in time and space. Numerical solutions  to optimal control of PDEs often base on computationally expensive iterative algorithms restricted to relatively small spatial domains and short time intervals. Clearly, the convergence to the target state sensitively depends on an appropriate initial guess for the control signal.

For traveling wave patterns, a general control task is position control aimed at guiding the pattern according to a given \textit{protocol of motion} (POM), i.e., moving it with desired velocity along a desired trajectory through a spatial domain. In some technical applications like catalytic reactors,  it is necessary to avoid the collision of high-temperature spots with the reactor walls or their pinning at heterogeneities of the catalyst's support to maintain operational safety \cite{Sheintuch2008}. Another example of open-loop position control is the enhancement of the $\mathrm{CO}_2$ production rate during the low-pressure catalytic oxidation of $\mathrm{CO}$ on $\mathrm{Pt}(110)$ single crystal surfaces by dragging of reaction pulses and fronts using a focused laser beam with speeds differing from their natural propagation velocity in the absence of control \cite{WolffPRL2003,qiao_enhancement_2008}. In a photosensitive Belousov-Zhabotinsky (BZ) medium, periodic variations of the applied light intensity forces a spiral wave tip to describe a wide range of hypocycloidal and epicycloidal trajectories \cite{steinbock1993control,Schrader1995}. In optical bistable media like dye-doped liquid crystals and Kerr cavities, interface dynamics can be controlled by spatially inhomogeneous forcing \cite{odent_optical_2016}. Position control of traveling wave patterns can be tackled by feedback control as well. For example, the spiral wave core in a photosensitive BZ medium was steered around obstacles using feedback signals obtained from wave activity measured at a point detector, from tangential crossing of wavefronts with detector lines, or a spatially extended control domain \cite{Zykov2003,Zykov2004,Schlesner2008book}. Two feedback loops were used to stabilize and guide unstable traveling wave segments along pre-given trajectories \cite{sakurai2002}. Furthermore, feedback-mediated control loops were employed to stabilize plane waves undergoing transversal instabilities \cite{STotz2018}.

Recently, we proposed an open-loop control that acts solely via the Goldstone modes of wave patterns \cite{loeber_engel2014} and provides analytical expressions for the amplitude of the control signal to be applied; it is coined \textit{Goldstone mode control}. We demonstrated that this control is able to accelerate or decelerate 1D traveling front and pulse solutions to RD equations \cite{loeber_engel2014,buch_eng_kamm_tro2013,LoeberBook2014} without changing their spatial profile. The stability of the control loop with respect to small changes in the initial conditions was discussed in \cite{lober2014stability}. Goldstone mode control also applies to move the core of a spiral wave at desired velocity along a pre-given trajectory through a 2D spatial domain, or to shape iso-concentration lines of 2D traveling pulses \cite{lober2014shaping}. Interestingly enough, the control turned out to be equivalent to the solution of an appropriately formulated optimal control problem \cite{loeber_engel2014,Ryll2016}.

In this paper, we extend Goldstone mode control to spatially localized moving spots. We introduce a three-component RD model supporting stable traveling spot solutions in \secref{sec:model} and derive analytical expressions for position and orientation control of traveling spots in the fully-actuated case in \secref{subsec:anapos}. The corresponding optimal control problem with an objective functional involving a Tikhonov regularization term is formulated explicitly in \secref{subsec:optpos}. Here, we discuss the relation between Goldstone mode control derived in \secref{subsec:anapos} and the solutions to the optimal control problem. In section \ref{sec:examples}, after a brief description of the numerical methods being used, we discuss examples for fully-actuated position and orientation control of spots in subsection \ref{subsec:position_control} and \ref{sec:orientation_control}, respectively, as well as for under-actuated position control by a single control signal \ref{subsec:singcontrol}. Finally, we conclude the results in \secref{sec:conclusion}.

\begin{figure}
\centering
\includegraphics[width=0.7\linewidth]{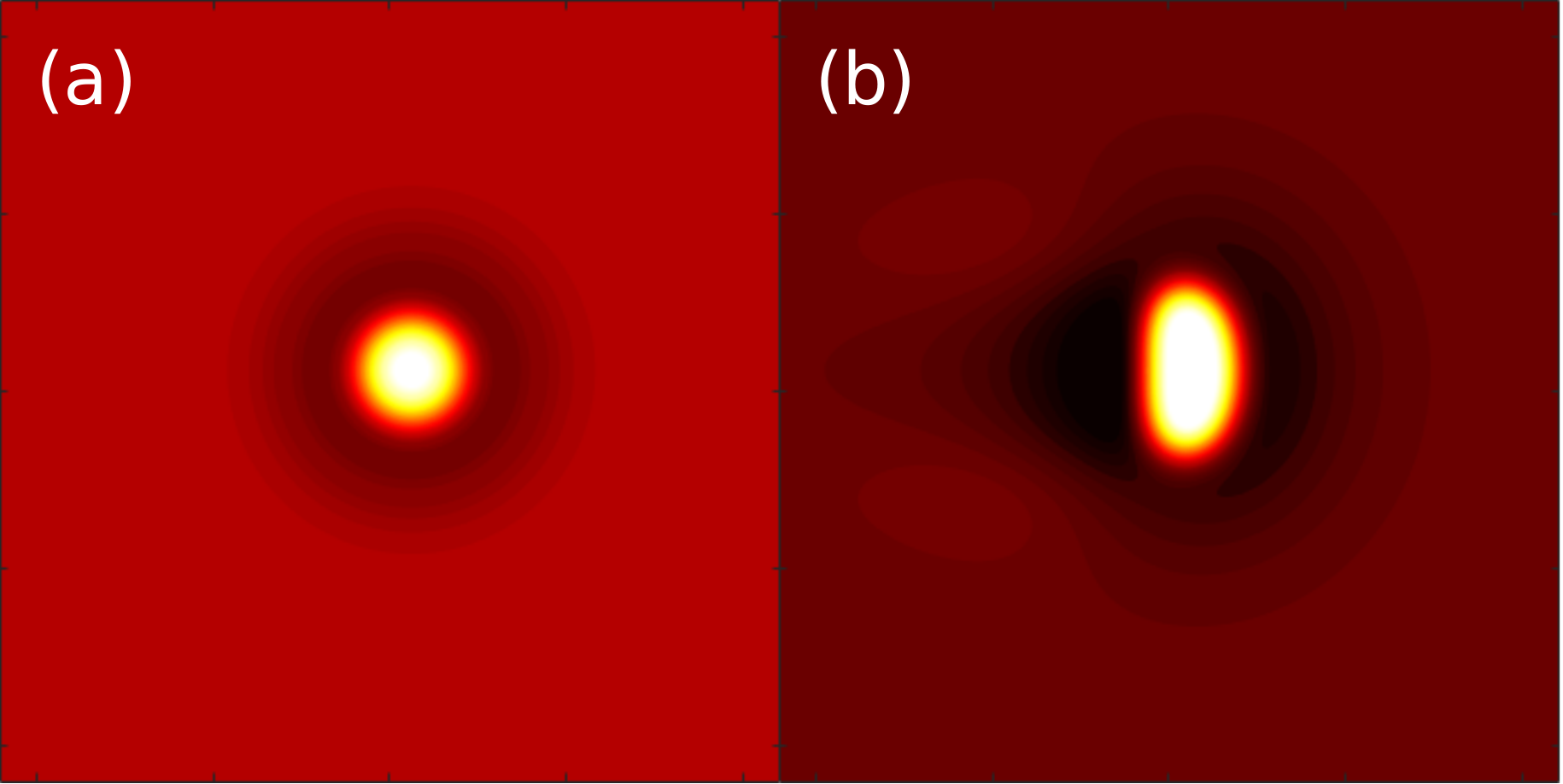}
\caption{Activator distribution $u$ of a rotational symmetric, resting  (a) and an axis-symmetric, traveling spot solution (b) to \eq{eq:spot_model}. Parameters $\kappa_1=-5.0$, $\tau=2$ in (a) and $\kappa_1=-6.92$, $\tau=48$ in (b); remaining parameters are taken from set $1$ in \tabref{tab:simu_parameter}. In both panels, identical domain size and limits for the colormap are used.} 
\label{fig:spot_profiles}
\end{figure}

\section{Three-component spot model} \label{sec:model}

Throughout this work, we consider the following three-component RD system exhibiting immobile and traveling stable spot solutions in $2$D \cite{purwins_dissipative_2005,doelman_pulse_2009,van_heijster_pulse_2008}

\begin{subequations} \label{eq:spot_model}
\begin{align}
% \begin{split}
\partial_t u(\V{r},t) =& D_u \Delta u + \kappa_2\,u -u^3- \kappa_3 v -\kappa_4 w + \kappa_1, \label{eq:spot_modela}\\
\tau \partial_t v(\V{r},t) =& D_v \Delta v + u -v, \label{eq:spot_modelb}\\
\theta \partial_t w(\V{r},t) =& D_w \Delta w + u -w, \quad \V{r} \in \Omega. \label{eq:spot_modelc}
% \end{split}
\end{align} 
\end{subequations}
Here, $\Delta=\partial_x^2+\partial_y^2$ represents the Laplacian in Cartesian coordinates, $\V{r}$ is the position vector in the spatial domain $\Omega$, $\V{r}=(x,y)^T \in \Omega \subset \R^2$, and $t$ indicates time. $D_u,D_v$, and $D_w$ denote the diffusion coefficients of components $u,v$, and $w$ while $\tau$ and $\theta$ set the time scales for the $v$ and $w$ kinetics, respectively. Beside spots, the model \eq{eq:spot_model} is capable to support peanut patterns \cite{nishiura2011}, breathing solitons \cite{gurevich_breathing_2006}, and jumping oscillons \cite{yang_jumping_2006}, for example.

\begin{figure}[th!]
\centering
\includegraphics*[width=1\textwidth]{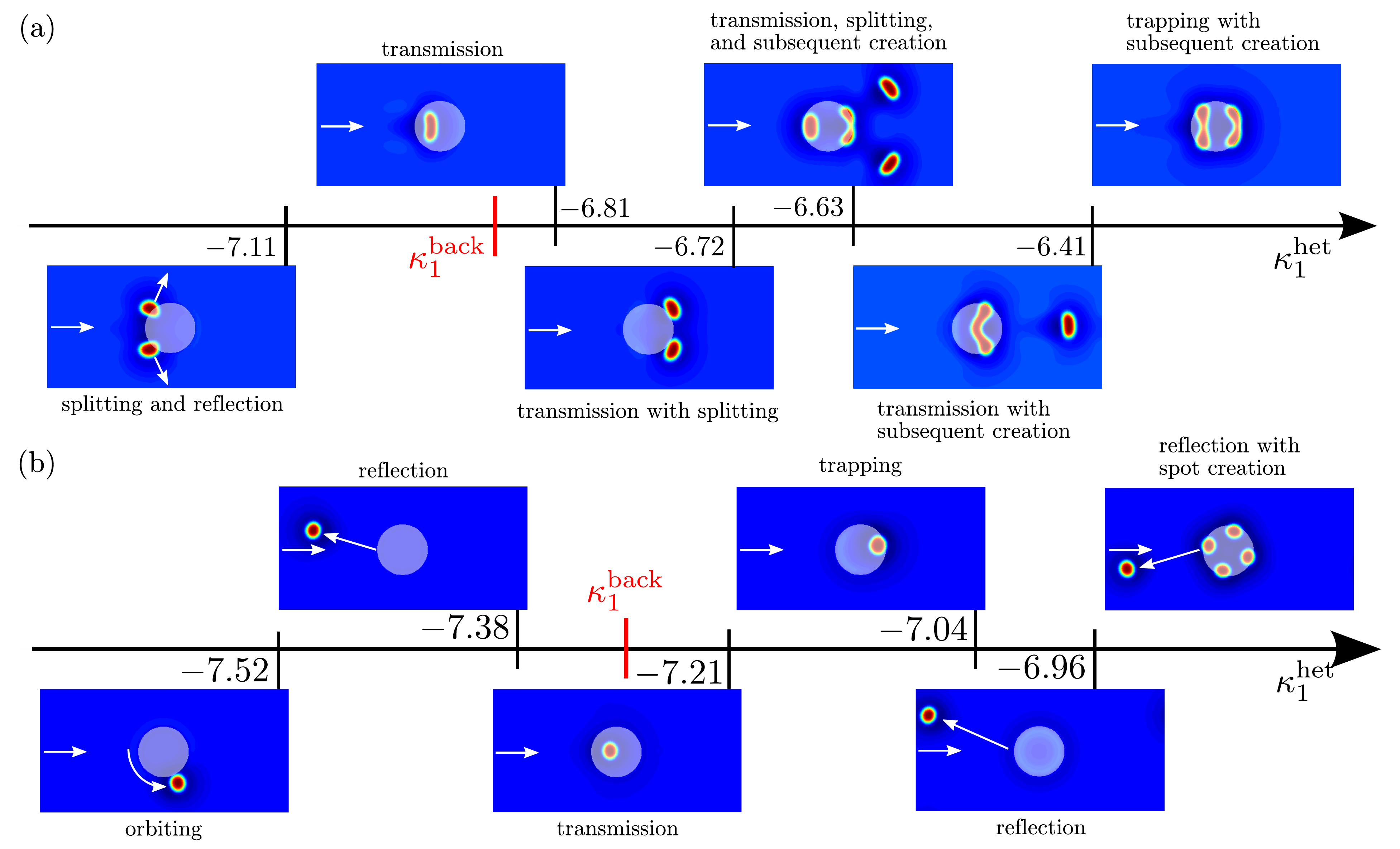}
\caption[Responds of traveling spots interacting with a circular defect] {Traveling spot interacting with a circular defect shown in light blue. (a) Splitting and different regimes of transmission or nucleation of new spots [SI video1]. Parameter set $1$ in \tabref{tab:simu_parameter} with  $\kappa_1^\mathrm{back} = -6.92$. (b) Trapping, reflection, transmission, and nucleation of new spots. Parameter set 2 in \tabref{tab:simu_parameter} with $\kappa_1^\mathrm{back} = -7.30$.	Shown are snapshots of the activator distribution obtained by numerical simulations of  \eq{eq:spot_model} on a rectangular spatial domain of size $\Omega=[0, 1) \times [-0.25 ,0.25)$ with periodic boundary conditions. Simulations were performed using ETD2; for details please see supplementary information S1.}
\label{fig:interaction_purwins}
\end{figure}

The three-component RD system \eq{eq:spot_model} was first introduced by Purwins and co-workers to model the dynamics of current filaments in planar gas-discharge \cite{purwins_dissipative_2005}.
In this context, activator $u$ and inhibitor $v$ represent the current density and the voltage drop over the high-ohmic electrode, respectively. The second inhibitor $w$ is linked to the surface charge, and the additive bifurcation parameter $\kappa_1$ is related to the supply voltage. Replacing the constant additive parameter in \eq{eq:spot_model} by a space-dependent quantity $\kappa_1(\V{r})$ breaks the translation and the rotation Euclidean symmetries of the equations. The interaction of traveling spots with different types of parameter heterogeneities in $1$D and $2$D has been studied in detail by many authors, see \cite{nishiura2011} and references therein. Penetration, rebound, annihilation, oscillation, as well as stationary or oscillatory pinning of spots were observed. Figure \ref{fig:interaction_purwins} illustrates different outcomes of the interaction between a traveling spot and a localized circular defect formed by a finite jump $\delta k_1= \kappa_1^\mathrm{het}-\kappa_1^\mathrm{back}$ from a background value $\kappa_1^\mathrm{back}$ and a higher value $\kappa_1^\mathrm{het}$ inside the circular heterogeneity. Additionally, other scenarios of spot-defect interaction have been found like repeated creation of spots inside the heterogeneity as well as spots orbiting both inside and outside of the defect's boundary.
If the activator describes the temperature in a catalytic packed-bed reactor, resting \emph{hot spots} \cite{marwaha_hot_2003} or those pinned to local heterogeneities can damage the catalyst support. In particular, collision of \emph{hot spots} with the reactor walls must be prevented for safety reasons. Consequently, guidance of a traveling spot with given velocity along a desired trajectory through a bounded spatial domain might be particular challenge in chemical engineering applications.

\section{Controlling position and orientation of traveling spots}\label{S1}
\subsection{Analytical expression for control amplitudes in position control} \label{subsec:anapos}
\noindent Let us consider a controlled RD system according to
\begin{subequations} \label{eq:contRDSwithCond}
\begin{equation}
\partial_t \V{U}(\V{r},t) -  \mathbb{D} \Delta \V{U}(\V{r},t) - \V{R}(\V{U}(\V{r},t)) = \mathbb{B}\V{f}(\V{r},t) \label{eq:contRDS}.
\end{equation}
Here, $\V{U}(\V{r},t)=(u_1(\V{r},t),\dots,u_n(\V{r},t))^T$ is the vector of $n\in\mathbb{N}$ state components defined in the two-dimensional spatial domain $\Omega \subset \R^2$ with $\V{r}=(x,y)^T$.  Assuming an isotropic medium, the $n\times n$ matrix of diffusion coefficients $\mathbb{D}$ is diagonal and constant, $\mathbb{D}=\mathrm{diag}(D_1,\ldots,D_n)$. The vector $\V{R(\V{U})}=(R_1(\V{U}),\dots,R_n(\V{U}))^T$ describes the reaction kinetics of the components. In general,  $R_i(\V{U})$ are nonlinear functions of the state. For the RD system \eqs{eq:spot_model},  $\V{U}$, $\mathbb{D}$, and $\V{R}$ are given by $\V{U}=(u,v,w)^T$, $\mathbb{D}=\mathrm{diag}(D_u,D_v/\tau,D_w/\theta)$, and $\V{R}=\bracket{\kappa_2\,u -u^3 - \kappa_3 v -\kappa_4 w + \kappa_1,(u - v)/\tau,(u - w)/\theta}^T$, respectively. 
Equation \eqref{eq:contRDS} must be supplemented with an initial condition 
\begin{equation}
\V{U}(\V{r},t_0)=\V{U}_0(\V{r}), \label{eq:contRDSIC}
\end{equation}
and appropriate boundary conditions.  We consider a rectangular domain $\Omega=(x_a, x_b] \times (y_a, y_b]$ with periodic boundary conditions such that $\V{U}$ as well as its derivatives in the direction normal to the boundary are periodic,
\begin{equation}
\begin{aligned}
\V{U}(x_a,y,t) =&\, \V{U}(x_b,y,t), &\quad\quad  \p{^l}{^l x}\V{U}(x_a,y,t) =&\, \p{^l}{^l x}\V{U}(x_b,y,t), \\
\V{U}(x,y_a,t) =&\, \V{U}(x,y_b,t), &\quad\quad  \p{^l}{^l y}\V{U}(x,y_a,t) =&\, \p{^l}{^l y}\V{U}(x,y_b,t),\, l\geq 1.
\end{aligned} \label{eq:contRDSBC}
\end{equation} 
\end{subequations}
The space-time dependent control signals $\V{f}(\V{r},t)=(f_1(\V{r},t),\dots,f_m(\V{r},t))^T$, $m\in\mathbb{N}$, on the right hand side of \eq{eq:contRDS} are assumed to act for all times $t$ everywhere within $\Omega$. 

The constant $n\times m$ matrix $\mathbb{B}$ determines which components are directly affected by the control signals. A system with strictly less independent control signals than components, $m<n$, is \emph{underactuated}. For $m=n$ and $\mathbb{B}$ invertible, the system is \emph{fully actuated}. In what follows, we focus on fully actuated systems and set $\mathbb{B}$ equal to the identity matrix $\mathds{1}$. The limiting case of single component control, i.e., $\mathbb{B}\V{f}(\V{r},t)\propto (f_1(\V{r},t),0,\dots,0)^T$, we consider in subsection \ref{subsec:singcontrol}.

The partial differential equations \eq{eq:contRDS} describe the evolution of the components $\V{U}(\V{r},t)$ in the presence of spatio-temporal perturbations $\V{f}(\V{r},t)$ that break the translation and rotation invariance of the unperturbed equations. In this interpretation, the response of the unperturbed solution to a given small input $\V{f}$ can be calculated perturbatively, see Ref. \cite{mikhailov1983stochastic,Nishiura2005,Biktashev2015}, and the SI. 

In this paper, following \cite{loeber_engel2014}, for given desired spot dynamics, we perceive \eq{eq:contRDS} as a conditional equation for the perturbations which now are considered as control inputs. The goal of the control $\V{f}$ is to enforce a state $\V{U}$ to follow a given \textit{desired distribution} $\V{U}_d(\V{r},t)=\bracket{u_{1,d}(\V{r},t),\ldots,u_{n,d}(\V{r},t)}^T$ as closely as possible everywhere in the spatial domain $\Omega $  and for all times $0 \leq t\leq T$. We call a desired distribution $\V{U}_d$ \emph{exactly realizable} if there exists a control $\V{f}$ such that the controlled state $\V{U}$ equals $\V{U}_d$ everywhere in the space-time cylinder $Q = \Omega \times [0,T]$.

Inserting $\V{U}_d$ for $\V{U}$ in \eq{eq:contRDS} yields for the control 
\begin{equation}
\V{f}(\V{r},t)=\mathbb{B}^{-1} \lbrace \displaystyle \partial_t \V{U}_d(\V{r},t) - \mathbb{D}\Delta \V{U}_d(\V{r},t) - \V{R}(\V{U}_d(\V{r},t))\rbrace. \label{eq:fins}
\end{equation}
For $\V{U}_d$ to be exactly realizable, three more conditions must be satisfied: First, the initial condition for the controlled state, \eq{eq:contRDSIC}, must coincide with the initial state of the desired distribution, $\V{U}(\V{r},0) = \V{U}_d(\V{r},0)$. Second, all boundary conditions for the desired distribution $\V{U}_d$ have to comply with the boundary conditions for $\V{U}$, \eq{eq:contRDSBC}. Third, $\V{U}_d$ must be sufficiently smooth in the space-time cylinder $Q = \Omega \times [0,T]$ such that the derivatives $\partial_t \V{U}_d$ and $\Delta \V{U}_d$ are continuous.
%
%===========================================Spot control
%

\begin{figure}
\centering
\includegraphics[width=0.8\linewidth]{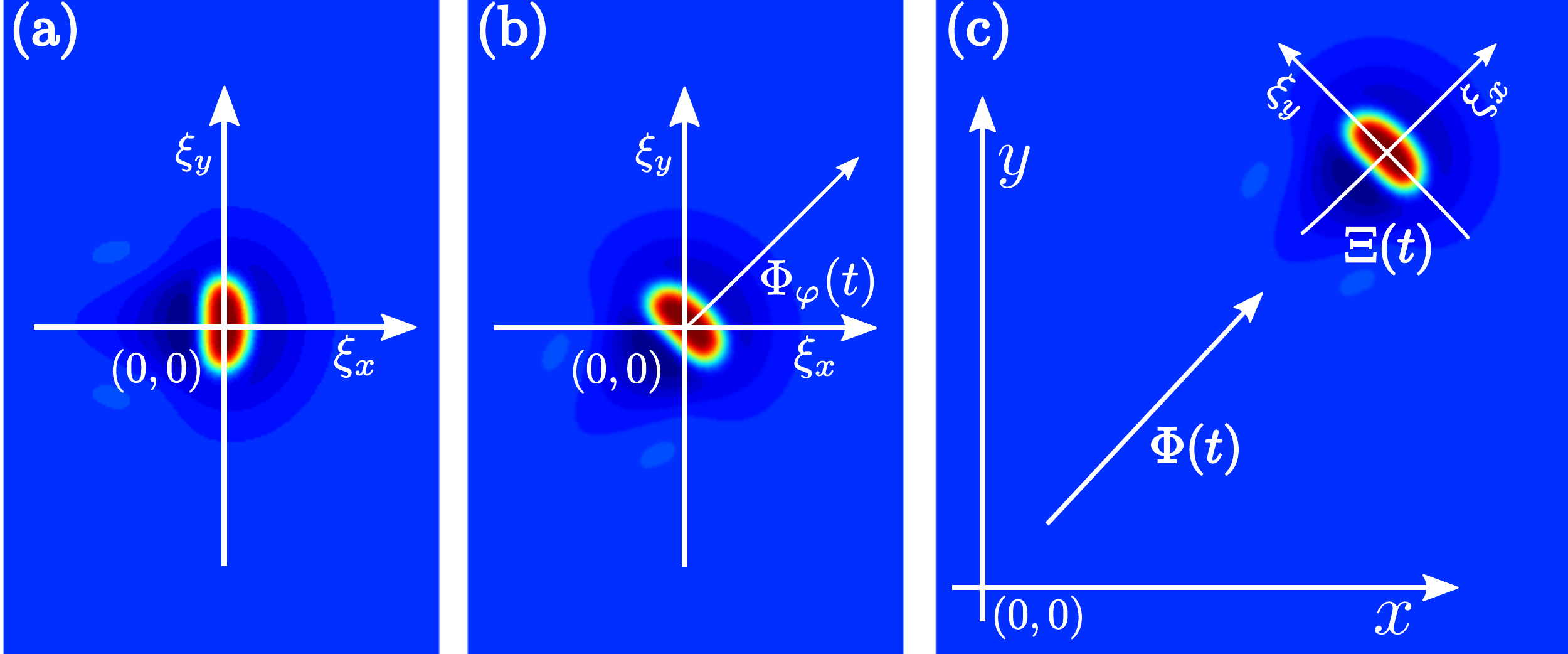}
\caption{Position and orientation of the spot $\V{U}_d\bracket{\V{r},t}$ as defined in \eq{eq:PositionControlDesiredTrajectory}. (a) Wave profile of the activator distribution, $U_c(\grb{\xi})$, centered in the co-moving and co-rotating frame of reference at $\grb{\xi}=(0,0)^T$. (b) Counter-clockwise rotation of $U_c(\grb{\xi})$ according to the desired orientation $\Phi_\varphi(t)$. (c) Shift of the rotated solution in virtue of the translational protocol of motion $\grb{\Phi}(t)=(\Phi_x(t),\Phi_y(t))^T$.} 
\label{fig:explanation_shift}
\end{figure}

Next, we formulate the control goal for spot solutions to the uncontrolled RD equations\eq{eq:contRDSwithCond}. These solutions propagate with constant velocity $\V{v}_0=(v_0^x,v_0^y)^T$ and wave profile $\V{U}_c$ through the spatial domain. In a co-moving frame of reference, $\grb{\xi}=(\xi_x,\xi_y)^T \equiv\V{r}-\V{v}_0t$, $\V{U}_c$ satisfies the equation
\begin{align}
\V{0}=&\mathbb{D} \Delta_{\grb{\xi}} \V{U}_c\bracket{\grb{\xi}} + \V{v}_0\cdot \nabla_{\grb{\xi}} \V{U}_c\bracket{\grb{\xi}} + \V{R}(\V{U}_c\bracket{\grb{\xi}}), \label{eq:ProfileEquation}
\end{align}
where, $\nabla_{\grb{\xi}}=(\partial_{\xi_x},\partial_{\xi_y})^T$ and $\Delta_{\grb{\xi}}=\partial_{\xi_x}^2+\partial_{\xi_y}^2$ denote the component-wise gradient and Laplacian, respectively. We emphasize that resting localized spots, $\V{v}_0=\V{0}$, are rotationally-symmetric solutions while traveling localized spots are axis-symmetric with the symmetry axis directed tangentially to the trajectory of motion, cf.\@ \figref{fig:spot_profiles}(a) and (b), respectively. We characterize the current position of a spot by the $x$- and $y$-coordinates of the maximum value of the activator concentration along its symmetry axis at a given time, $\grb{\Phi}(t)=(\Phi_x(t),\Phi_y(t))^T$, and its orientation by the angle $\Phi_\varphi(t)$  between the spot's symmetry axis and the $x$-axis, compare figure \ref{fig:explanation_shift}.

A distribution following a prescribed POM $\grb{\Xi}(t)=(\grb{\Phi}(t),\Phi_\varphi(t))^T$, while simultaneously preserving the profile of the uncontrolled spot $\V{U}_c$, reads
\begin{equation}
\V{U}_d(\V{r},t)=\V{U}_c(\mathbb{A}(-\Phi_\varphi(t))\bracket{\V{r}-\grb{\Phi}(t)}). \label{eq:PositionControlDesiredTrajectory}
\end{equation}
Here, $\mathbb{A}(\alpha)=\left[\,\cos(\alpha),-\sin(\alpha);\, \sin(\alpha),\cos(\alpha)\, \right]$ is the clockwise rotation matrix in $2$D. For the desired distribution \eq{eq:PositionControlDesiredTrajectory} to be exactly realizable, the initial condition must be a spot solution of the form $\V{U}(\V{r},t_0)=\V{U}_c(\mathbb{A}(-\phi_0)\bracket{\V{r}-\V{r}_0})$, which yields for the initial values of the POM $\grb{\Phi}(t_0)=\V{r}_0$ and $\Phi_\varphi(t_0)=\phi_0$, respectively. Inserting the desired distribution \eq{eq:PositionControlDesiredTrajectory} into the general control solution \eq{eq:fins} leads to

\begin{align}
\V{f}(\V{r},t)=\,& \left[-\bracket{\mathbb{A}_z(-\Phi_\varphi(t))\, \dot{\grb{\Xi}}(t)} \cdot \tilde{\nabla}_{\grb{\xi}} \V{U}_c(\grb{\xi}) -\mathbb{D} \Delta_{\grb{\xi}} \V{U}_c(\grb{\xi}) - \V{R}(\V{U}_c(\grb{\xi}))\right]_{\grb{\xi}=\bar{\grb{\xi}}(t)},
\end{align}
with $\bar{\grb{\xi}}(t) = \mathbb{A}(-\Phi_\varphi(t))\bracket{\V{r}-\grb{\Phi}(t)}$. For the sake of a compact notation, we introduced the differential operator $\tilde{\nabla}_{\grb{\xi}}=(\partial_{\xi_x},\partial_{\xi_y}, \partial_{\varphi})^T$ with the angular derivative $\partial_{\varphi} = -\xi_y \partial_{\xi_x} + \xi_x \partial_{\xi_y}$. The dot denotes the derivative with respect to time $t$, and $\mathbb{A}_z(\alpha)$ is the clockwise rotation matrix around the $z$-axes in $3$D, $\mathbb{A}_z(\alpha)=\mathrm{diag}(\mathbb{A}(\alpha),1)$. Using equation \eq{eq:ProfileEquation} for the uncontrolled spot profile, we end up with the expression
\begin{align}\label{eq:fana}
\V{f}_\mathrm{Gold}(\V{r},t) =\,\left[\bracket{ \Vthree{v_0^x}{v_0^y}{0} -\mathbb{A}_z\bracket{-\Phi_\varphi(t)} \Vthree{\dot{\Phi}_x(t)}{\dot{\Phi}_y(t)}{\dot{\Phi}_\varphi(t)} }\cdot \tilde{\nabla}_{\grb{\xi}}\right]\V{U}_c(\grb{\xi})\Bigg|_{\grb{\xi}=\bar{\grb{\xi}}(t)}
\end{align}
for our Goldstone mode control. 

Remarkably, any reference to the nonlinear functions $\V{R}$ drops out from the result \eq{eq:fana}. This is of great advantage in all applications where the details of the underlying reaction kinetics $\V{R}$ are largely unknown or difficult to identify. Once propagation velocity $\V{v}_0$ and wave profile $\V{U}_c$ of the uncontrolled spot are measured with an accuracy sufficient to calculate the Goldstone modes $\partial_{\xi_x}\V{U}_c$, $\partial_{\xi_y}\V{U}_c$, and $\partial_{\varphi}\V{U}_c$, the control signals can be computed in advance for the complete time interval $[0,T]$. Consequently,
in contrast to feedback control, a continuous recording of the system is not required.

One notices that \eqref{eq:fana} equals the sum of Goldstone modes with time-dependent prefactors, $\V{f}_\mathrm{Gold}(\V{r},t)=\,P_1(t)\, \partial_{\xi_x}\V{U}_c(\grb{\xi}) + P_2(t) \,\partial_{\xi_y}\V{U}_c(\grb{\xi}) +P_3(t)\, \partial_{\varphi}\V{U}_c(\grb{\xi})$. The Goldstone modes are the right eigenvectors to the linear stability operator $\mathcal{L}$ of \eq{eq:ProfileEquation} 
\begin{align}
\mathcal{L}=\,\mathbb{D} \Delta_{\grb{\xi}} + \V{v}_0\cdot \nabla_{\grb{\xi}}+\mathcal{D}\V{R}(\V{U}_c\bracket{\grb{\xi}}), \label{eqApp:stabOp}
\end{align}
to the eigenvalue zero. They are associated with  the translational and rotational invariance of equation \eqref{eq:contRDS} in $\R^2$ for $\V{f}(\V{r},t)=0$. Clearly, the prefactors' magnitudes are proportional to the difference between the intrinsic velocity, $\V{v}_0$, and the current prescribed spot velocity projected onto the $x$- and $y$-axes. If the prescribed POM $\grb{\Xi}(t)$ coincides with the spot's natural motion, then all prefactors vanish identically and $\V{f}_\mathrm{Gold}$ disappears everywhere in $Q$.
Importantly, the control signal is localized around the spot position and vanishes far from it because the spatial derivatives of its profile decay sufficiently fast, $\lim_{\norm{\grb{\xi}}\to \infty} \nabla_{\grb{\xi}} \V{U}_c=\V{0}$.

Alongside with these advantages, limitations in the applicability of Goldstone mode control \eqref{eq:fana} exist as well. For instance, the magnitude of the applied control may locally attain values that are unfeasible to realize physically because $\V{f}_\mathrm{Gold}$ is proportional to the slope of the controlled wave profile $\V{U}_c$. The stability of the control scheme depends sensitively on how precise the Goldstone modes can be calculated. Further, the complete spatial domain $\Omega$ accessible by the spot has to be available for the control as well. Additionally, as already mentioned above, $\V{f}_\mathrm{Gold}$ cannot be applied to desired trajectories $\V{U}_d$ which do not comply with initial as well as boundary conditions or which are non-smooth. While all these cases cannot be treated within the analytical approach proposed here, optimal control can deal with many of these complications. 

%=================================================
%=================================================Optimal control
%
\subsection{Optimal control} \label{subsec:optpos}

An optimal control minimizes a so-called \textit{objective functional} $J$ defined as a non-negative tracking-type functional
\begin{align} \label{eq:functopt}
J(\V{U},\V{f})=&\,\frac{1}{2}\sum_{i=1}^3\left[\intQ{ \bracket{u_i-u_{i,d}}^2 }+ \nu\intQ{f_{i}^{2}}\right]\!.
\end{align}
$\V{U}$ satisfies the controlled state equation associated to $\V{f}$ with respect to given initial and boundary condition, cf.\@ equations \eq{eq:contRDSwithCond}. The first term appearing in $J$ measures the distance between the actual and the desired solution $\V{U}$ and $\V{U}_d$ up to the terminal time $T$ in an $L^2(Q)$-sense. In the second, so-called Tikhonov regularization term, a small but finite, positive value $\nu$ guarantees the existence of an optimal control $\V{f}_\mathrm{opt}$ that minimizes the objective functional $J$ \eq{eq:functopt} for $\Omega \subset \R^q, q = 1,2,3$, see Ref.~\cite{Casas2018}.

For exactly realizable desired states, $ \V{U}=\V{U}_d$, the solution to \eq{eq:fins} equals the solution to the unregularized optimal control problem for $\nu=0$. If $\V{U}_d$ is not exactly realizable, the controlled state $\V{U}$ must be obtained as part of the solution to the optimal control problem. The minimization of $J$ must be performed with respect to state $\V{U}$ and control $\V{f}$. Expressing $\V{U}$ in terms of $\V{S}(\V{f})$, where $\V{S} : \V{f} \mapsto \V{U}$ is the solution operator to \eq{eq:contRDSwithCond} in $Q$, justifies the definition of a reduced objective functional $J(\V{f}) := J(\V{S}(\V{f}),\V{f})$. In order to minimize $J(\V{f})$, its first directional derivative with respect to $\V{f}$ has to equal zero in all directions $\V{h}$; yielding the necessary optimality conditions
\begin{equation} \label{eq:NecOptCond}
\left[\intQ{\left(\left(\V{S}(\V{f}_\mathrm{opt})\right) - \V{U}_{d}\right)\cdot \left(\V{S}^\prime(\V{f}_\mathrm{opt})\V{h}\right) \,} + \nu\intQ{\V{f}_\mathrm{opt}\cdot \V{h}\, }\right] = 0\quad \forall \, \V{h}.
\end{equation}
The state $\V{U}$ is constrained to satisfy the controlled state equation together with given initial and boundary conditions, cf.\@ equations \eq{eq:contRDSwithCond}. Similar as in ordinary minimization problems, a constrained minimization can be transformed to an unconstrained one by introducing Lagrange multipliers $\V{P}(\V{r},t) = (p_1(\V{r},t),\ldots,p_n(\V{r},t))^T$, also called the \textit{adjoint state}. By means of the latter, \eq{eq:NecOptCond} can be reformulated
\begin{equation}  \label{eq:NecOptCond2}
\intQ{\left(\V{P} + \nu \, \V{f}_\mathrm{opt}\right)\cdot \V{h}\, } = 0 \quad \forall \, \V{h},
\end{equation}
whereby the adjoint state is the solution of the \textit{adjoint equation}
\begin{equation}\label{eq:adj}
- \partial_t \V{P}(\V{r},t) - \mathbb{D}\Delta\V{P}(\V{r},t)
-\mathcal{D}\V{R}^T(\V{U}_\mathrm{opt}(\V{r},t))\V{P}(\V{r},t) = \V{U}_\mathrm{opt} - \V{U}_d \quad \mbox{in}\quad Q,
\end{equation}
subject to terminal condition $\V{P}(\cdot,T)=\V{0}$ in $\Omega$ and periodic boundary conditions in $\partial\Omega$. Here, $\mathcal{D}\V{R}^T$ denotes the transposed Jacobian matrix of $\V{R}$ with respect to $\V{U}$. It is rather obvious that the condition \eqref{eq:NecOptCond2} is equivalent to the condition 
\begin{equation} \label{eq:gradient_condition}
\V{P} + \nu \, \V{f}_\mathrm{opt} = \mathbf{0}.
\end{equation}
This is nothing more than the well-known condition that, in a minimum, the gradient of the function to be minimized is zero. 

Due to the mixed initial and terminal conditions for $\V{U}$ and $\V{P}$ it is rarely possible to find numerical solutions to optimal control by a direct integration method. To reduce numerical costs, we employ Model Predictive Control and divide our optimal control problem in subproblems with a $4$ time-step small time-horizon \cite{Ryll2016}. Thereby, each subproblem is solved with a gradient-type method. Details on the iteration scheme are discussed in the supplementary information (SI), paragraph S1.

%===========================================================
%===========================================================Examples
\section{Examples} \label{sec:examples}
In the following, we discuss three examples for position control of traveling spot solutions to the three-component RD model \eq{eq:spot_model}. Mainly, we compare Goldstone mode control $\V{f}_\mathrm{Gold}$ with optimal control $\V{f}_\mathrm{opt}$. 
If not stated otherwise, the state equation \eq{eq:contRDS} and the adjoint equation \eq{eq:adj} are solved on a squared domain \mbox{$\Omega = (-0.5,0.5]\times (-0.5,0.5]$} with periodic boundary conditions \eqref{eq:contRDSBC} both in $x$ and $y$. 
The domain size is sufficiently large to avoid self-interaction of the spots in the periodic simulation domain. Without loss of generality, we fix the spots' direction of motion to coincide with the $x$-axis, i.e., $v_0^x\neq 0$ and $v_0^y=0$. 
Any numerical simulation of equation \eqref{eq:contRDS} is initialized with the profile $\V{U}_c$ of the uncontrolled spot. This profile and the corresponding natural velocity $\V{v}_0$ are obtained by solving the nonlinear eigenvalue problem  \eqref{eq:ProfileEquation} with adequate accuracy. Further details on the used numerical methods, the spatial and temporal resolution, and the initial conditions are presented in the SI, S1.

\begin{table}
\begin{indented}
\item[]\begin{tabular}{@{}|l|c|c|c|c|c|c|}
	\hline
	& $D_u$ & $D_v$ & $D_w$ & $\tau$ & $\kappa_1$ & $v_0^x$ \\
	\hline
	set $1$ & $1.0\cdot 10^{-4}$ & $1.86 \cdot 10^{-4}$ & $9.6\cdot 10^{-3}$ & $48.0$ & -$6.92$ & $2.599\cdot 10^{-3}$ \\
	%			\hline
	%			set 1  \cite{bode_interaction_2002} & $1.5\cdot 10^{-4}$ & $1.86 \cdot 10^{-4}$ & $9.6\cdot 10^{-3}$ & $48.0$ & -$6.92$ & $3.318\cdot 10^{-3}$ \\
	\hline
	set $2$ \cite{nishiura2011} & $0.9\cdot 10^{-4}$ & $1.00 \cdot 10^{-3}$ & $1.0\cdot 10^{-2}$ & $40.0$ & -$7.30$ & $1.776\cdot 10^{-3}$  \\
	\hline
\end{tabular}
\end{indented}
\caption{\label{tab:simu_parameter} Parameter values used in the numerical simulations. The parameters $\theta=1,\,\kappa_2=2,$ $\kappa_3=1,$ and $\kappa_4=8.5$ are the same for set 1 and set 2.}
\end{table}

\subsection{Translational position control of spots} \label{subsec:position_control}

\begin{figure}
\centering
\includegraphics[width=\linewidth]{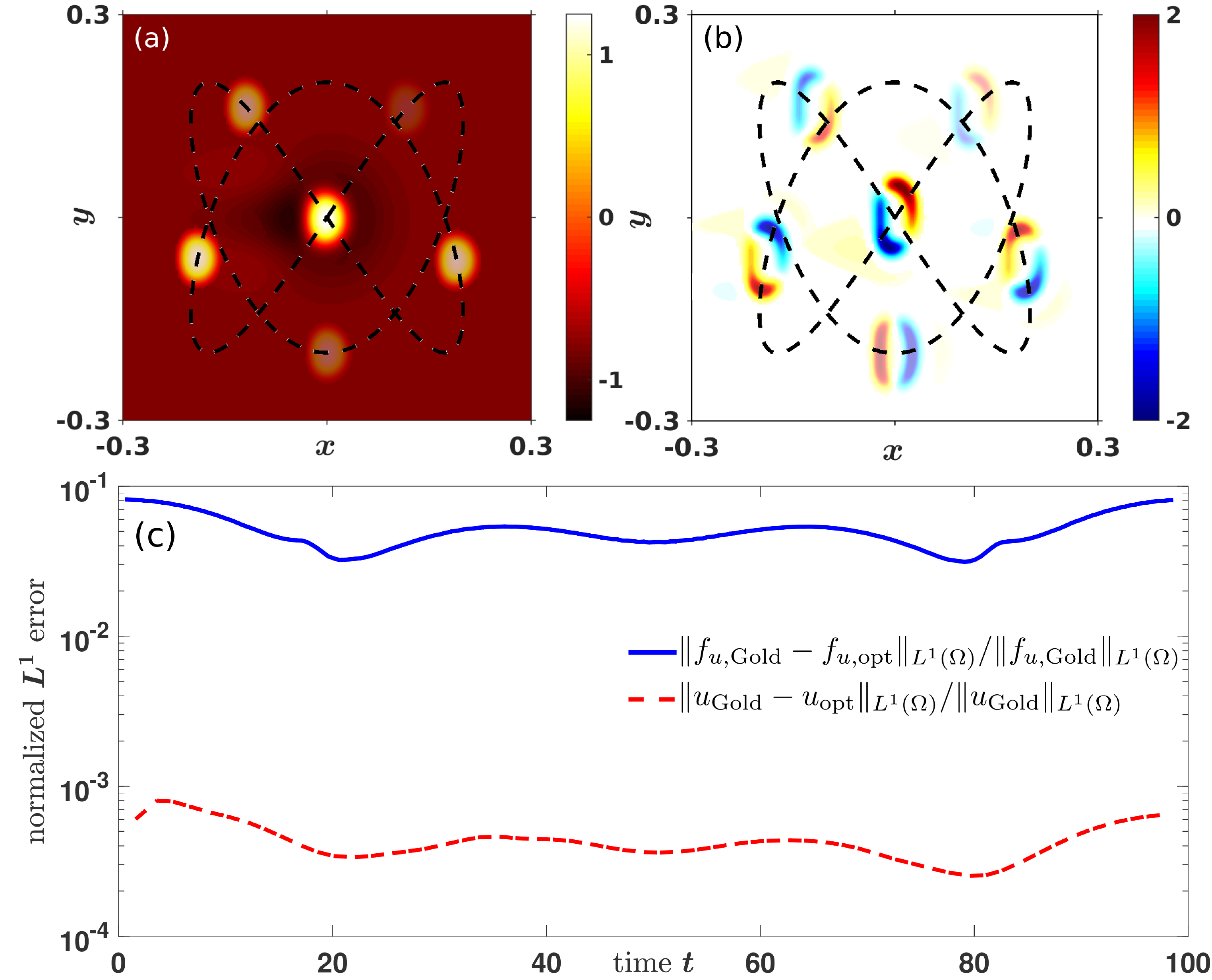}
\caption{Position control along the Lissajous curve \eq{eq:lissajous_curve}, see [SI\_video1]. (a) Snapshots of the activator distribution $u(\V{r},t)$ obtained from numerical simulation of \eq{eq:spot_model}-\eq{eq:contRDS} with control $\V{f}_\mathrm{Gold}$, \eq{eq:fana}, at time moments $t=\{10,50,90,130,170,200\}$. (b) Control $f_{u,\mathrm{Gold}}(\V{r},t)$ at the same instants of time. In (a) and (b), the dark dashed line indicates the Lissajous curve and the decreasing transparency marks consecutive time moments. 
(c) Temporal behavior of the relative $L^1(\Omega)$ error \eq{eq:def_l1norm} between $f_{u,\mathrm{Gold}}$ \eq{eq:fana}, and optimal activator control signals $f_{u,\mathrm{opt}}$ \eq{eq:functopt} during $t \in [0,T/2]$. 
We select set $1$ in \tabref{tab:simu_parameter} for the parameters to \eq{eq:spot_model} and set the Tikhonov parameter to $\nu=10^{-7}$.}
\label{fig:position_lissajous}
\end{figure}

In our first example, we aim to shift the spot's position along a Lissajous curve without controlling its orientation, i.e., the spot's symmetry axis is kept frozen to the $x$-axis. Thus, the POM \mbox{$\grb{\Xi}(t)=\bracket{\Phi_x(t),\Phi_y(t), \Phi_\varphi(t)}^T$} is given by
\begin{equation} \label{eq:lissajous_curve}
\Phi_x(t) =\, r\,\sin(4\pi\, t/T)),\quad \Phi_y(t) =\, r\,\sin(6\pi\,t/T)),\quad \mbox{and}\quad \Phi_\varphi(t) =\, 0,
\end{equation}
\noindent with radius $r=0.2$ and  protocol duration $T=200$. The video [SI\_video1] shows the complete dynamics of all three state components $\V{U}$ as well as $\V{f}_{\mathrm{Gold}}$ and $\V{f}_{\mathrm{opt}}$. In figure \ref{fig:position_lissajous}(a), we depict the time evolution of the activator distribution $u(\V{r},t)$ under the action of the control $\V{f}_\mathrm{Gold}$ shown in \figref{fig:position_lissajous}(b). One observes that the spot follows the desired trajectory indicated by the dashed line. The orientation of its symmetry axis remains fixed while the control signal realigns at any instants of time. Indeed, in the absence of orientation control, $\Phi_\varphi(t) =\, 0$,  $\V{f}_\mathrm{Gold}$ can be expressed by the projection of $\nabla \V{U}_c$ onto the tangential vector to the Lissajous curve $\V{T}(t)$,  
$\V{f}_\mathrm{Gold}(\V{r},t)=\left[\bracket{v_0^x-\dot{\Phi}_x(t)}\partial_{\xi_x} \V{U}_c\bracket{\grb{\xi}(t)}-\dot{\Phi}_y(t)\partial_{\xi_y} \V{U}_c\bracket{\grb{\xi}(t)}\right]\propto \V{T}(t)\cdot\nabla\, \V{U}_c$, with $\grb{\xi}(t)=\V{r}-\grb{\Phi}(t)$. Obviously, the control is localized at the current spot position $\grb{\Phi}(t)$ and vanishes far away from it. Despite that the average speed $\bar{v} = L_\mathrm{curve}/T \approx 6 v_0^x$ along the studied Lissajous curve \eq{eq:lissajous_curve} with arc length $L_\mathrm{curve}$ is almost five times larger than the propagation velocity of the uncontrolled spot, the magnitude of $f_{u,\mathrm{Gold}}$ is of the same order as the local reaction terms \eq{eq:spot_model}. The control signals applied to the inhibitors $v$ and $w$ are one and two magnitudes smaller [SI\_video1] than the activator's control, respectively.

On the scale of [SI\_video1], there is no distinguishable difference between $\V{f}_\mathrm{Gold}$ and $\V{f}_\mathrm{opt}$. Both are always localized close to the current spot position, and their magnitudes change proportional to $|\dot{\grb{\Phi}}(t)|$. For a quantitative comparison, we compute the relative errors between $f_{u,\mathrm{Gold}}$ and $f_{u,\mathrm{opt}}$ measured by the $L^1(\Omega)$-norm
\begin{equation} \label{eq:def_l1norm}
\norm{h(t)}_{\lao} = \intOm{|h\bracket{\V{r},t}|}.
\end{equation}
Here, $|h\bracket{\V{r},t}|$ indicates the absolute value of $h$ at position $\V{r}$ and time $t$.

In \figref{fig:position_lissajous}(c), we depict solely the normalized error for the first half of the protocol because it starts to repeat after $T/2$, $\Phi_y(t)=-\Phi_y(t+T/2)$. The relative error between $f_{u,\mathrm{Gold}}$ and optimal control $f_{u,\mathrm{opt}}$ (solid line) is satisfactory and ranges between $2\%$ and $8\%$. As reported in S1 of the SI, the limiting error is dominated by the time step chosen in the implicit Euler-scheme. Albeit the scheme is A-stable, the error at a specific time $t$ is of the order of $\mathcal{O}(dt)$. Consequently, we observe that $\norm{f_{u,\mathrm{Gold}}(t)-f_{u,\mathrm{opt}}(t)}_{\lao}$ is bounded from above by $dt$; $dt=0.1$ in the studied example. 

The dashed line in figure \ref{fig:position_lissajous}(c) shows the relative error between the activator distribution obtained by Goldstone mode control and the one calculated under optimal control, $\norm{u_{\mathrm{Gold}}(t)-u_{\mathrm{opt}}(t)}_{\lao}/\norm{u_{\mathrm{Gold}}(t)}_{\lao}$. At any time, this error is less than $10^{-3}$, i.e., both controlled states agree remarkably well, despite that  $\norm{f_{u,\mathrm{Gold}}(t)-f_{u,\mathrm{opt}}(t)}_{\lao}$ is of the order $10^{-1}$. Additionally, the relative errors between the desired distribution $\V{U}_d$ and the state solutions $\V{U}_\mathrm{Gold}$ and $\V{U}_\mathrm{opt}$ are less than $10^{-7}$ in both cases (not shown explicitly). This confirms that the Goldstone mode control \eq{eq:fana}, within numerical accuracy,  is indeed the solution to the corresponding unregularized optimal control problem. Similar conclusions had been obtained in our previous study of position control of front solutions in one spatial dimension, see \cite{Ryll2016}.

The gradient-type method, used to solve the optimal control problem, relies on an initial guess for the control signal. The closer the starting guess is to the final solution, the fewer iteration steps are necessary to converge for most established optimization methods. Starting every iteration with an initial zero control, it takes on average $\bar{n}_\mathrm{iter} \simeq 23$ iterations per time step for position control along the Lissajous curve \eq{eq:lissajous_curve}. Using the control solution of the previously solved subproblem as initial guess reduces the average number of iterations to $\bar{n}_\mathrm{iter} \simeq 14$. Taking advantage of the similarity between $\V{f}_\mathrm{Gold}$ and $\V{f}_\mathrm{opt}$, see \figref{fig:position_lissajous}(c), the computational costs reduce even further. The most substantial computational speed-up is obtained by initiating every optimization subproblem with \eqref{eq:fins}. Then, the iteration stops on average after the first step, $\bar{n}_\mathrm{iter} \simeq 1$.

%
%=========================================================
%=========================================================Stability of position control
%
\subsubsection{Stability of position control} \label{subsec:stability}
Any open-loop controls is sensitive against perturbations of the initial conditions, data uncertainty, or numerical roundoff errors. To test the stability of our Goldstone mode control for position control $\V{f}_\mathrm{Gold}$, we accelerate or decelerate a single spot from its initial, intrinsic velocity $\V{v}_0$ to a final velocity $\V{v}_1$ using a translational POM $\grb{\Xi}(t)=\bracket{\grb{\Phi},0}^T$ with velocity
\begin{align}
\dot{\Phi}_i(t) =&\begin{cases} v_0^i,& t< 0,\\
\frac{1}{2}\bracket{(v_0^i+v_1^i) +(v_0^i-v_1^i)\,\cos\bracket{\pi t/T_i}},& 0 \leq t \leq T_i,\\
v_1^i,& t > T_i,
\end{cases}
\label{eqs:sine_control}
\end{align}
for $i \in \{x,y\}$. Note that both the protocol's velocity $\dot{\grb{\Phi}}(t)$ and acceleration $\ddot{\grb{\Phi}}(t)$ are continuous functions within the interval $[0, T_i]$. $T_i$ denotes the duration of the protocol. The maximum acceleration $\pi \bracket{v_i^1-v_i^0}/(2 T_i)$ is proportional to the prescribed velocity difference $v_i^1-v_i^0$ and inversely proportional to $T_i$. 

\begin{figure}
\centering
\includegraphics[width=\linewidth]{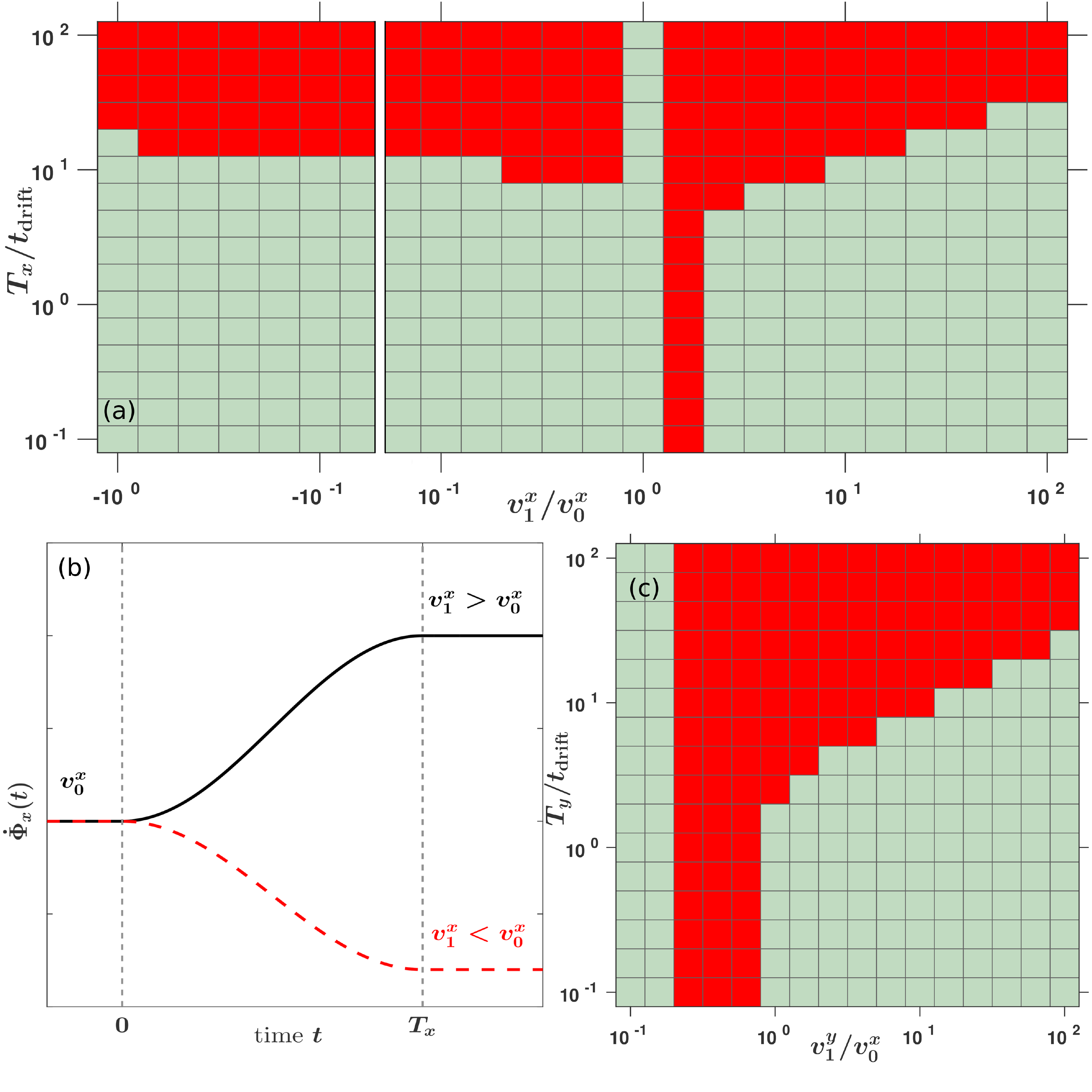}
\caption[Stability region for position control]{Numerically evaluated region of stability for position control $\V{f}_\mathrm{Gold}$. Stability (green regions) and instability (red regions) is demonstrated for an accelerating and decelerating POM \eq{eqs:sine_control} which changes the propagation velocity of a single spot from $\V{v}_0$ to the final value $\V{v}_1$ during $T_i,\,i=\lbrace x,y\rbrace$, see panel (b). In panel (a), the spot is exclusively accelerated or decelerated along its intrinsic direction of motion. In panel (c), the spot's propagation velocity perpendicular to the intrinsic one is controlled, $v_1^x=v_0^x$ and $v_1^y\neq 0$. The controlled spot dynamics \eq{eq:spot_model} is simulated on a $(-0.35, 0.35]\times (-0.35, 0.35]$ domain with periodic boundary conditions using ETD2, cf.\@ SI S1. The system parameters are taken from set $1$ in \tabref{tab:simu_parameter} and thus the drift time is given by $t_\mathrm{drift}= L/v_0^x\approx 273$.}
\label{fig:sketch_sinproc}
\end{figure}

A sketch of the protocol is depicted in \figref{fig:sketch_sinproc}(b). Since the proposed control scheme is an open-loop control, deviations between the current spot position $\grb{\Phi}_\mathrm{curr}(t)$ and the POM $\grb{\Phi}(t)$ will grow unbounded in time if the difference between them exceeds a critical value \cite{lober2014stability}. A specific protocol is called \textit{stable} and marked by green boxes in \figref{fig:sketch_sinproc} if and only if the Euclidean distance is bounded as $\norm{\grb{\Phi}_\mathrm{curr}(t)-\grb{\Phi}(t)} < L/2$ for all times $t \in [0,t_\mathrm{end}]$. Otherwise, it is called \textit{unstable} (red boxes). Note that a protocol is also considered to be unstable if the control leads to the nucleation of additional spots. In order to make the results comparable for different protocol durations, we adjust the terminal simulation time $t_\mathrm{end}$ according to $t_\mathrm{end}= \mathrm{max}\bracket{10\,t_\mathrm{drift},T_i+10\, L/|v_1^i|}$ with drift time $t_\mathrm{drift}=L/v_0^x$. We stress that all simulation results presented in \figref{fig:sketch_sinproc} have been computed for sufficiently long time intervals and do not alter upon an increase of the total simulation time.

Figure \ref{fig:sketch_sinproc}(a) depicts the numerically evaluated region of stable position control (green boxes) in $x$-direction as a function of the ratio of terminal spot velocity $v_1^x$ to the initial one $v_0^x$ and the ratio of the control duration $T_x$ to the drift time $t_\mathrm{drift}$. The translational POM in $y$-direction is set to zero, $\Phi_y(t)=0$. As expected, the numerical algorithm is stable in the absence of control, $v_1^x/v_0^x=1$. Further, it turns out that the control scheme is mostly stable for rapid, $T_x \ll t_\mathrm{drift}$, to moderately slow POMs, $T_x \lesssim 10\,t_\mathrm{drift}$, regardless of the velocity change, $| v_1^x - v_0^x|$. The stability regions exhibit an asymmetry with respect to the sign of the velocity change. Weakly accelerating protocols, $1 < v_1^x/v_0^x \lesssim 2$, are unstable (red colored region) while decelerating ones, $v_1^x < v_0^x$, are always stable for $T_x \lesssim 10\,t_\mathrm{drift}$. This finding is in agreement with \cite{lober2014stability}. The instability for $v_0^x < v_1^x \lesssim 2 v_0^x$ is caused by  an undesired rotation of the spot induced by numerical truncation errors. These accumulate during the simulation and eventually result in an asymmetric perturbation (with respect to $y$) acting on the spot pattern. Once the spot starts to rotate and eventually drifts away from the centerline $y=0$, the proposed open-loop control $\V{f}_\mathrm{Gold}$ can neither respond nor correct the undesired rotation. The impact of the numerical truncation error becomes more pronounced with growing protocol's duration $T_x$ and results in a broad unstable region for long protocols, $T_x > 10\,t_\mathrm{drift}$.

The situation changes if one aims to move the spot pattern perpendicular to its intrinsic direction of propagation, here in $y$-direction. In figure \ref{fig:sketch_sinproc}(c), we keep the motion in $x$ unchanged, $\Phi_x(t)=v_0^x\,t$, and accelerate the spot according to \eq{eqs:sine_control} along the $y$-direction. Because the controlled spot solution is symmetric with respect to the centerline $y=0$, position control in $y$ might be inherently unstable \cite{lober2014stability}. One notices immediately that regions with unstable position control are much larger compared to \figref{fig:sketch_sinproc}(a). Nevertheless, the control is stable for weak acceleration, $v_1^y \simeq 0.1 v_0^x$, independent of the protocol's duration. Increasing the terminal velocity further, Goldstone mode control starts to fail. Once a certain deviation between the current spots' position and the proposed POM is attained, the pattern cannot follow the applied control anymore and starts to move freely. 
With further growing terminal velocity $v_1^y$, the control's magnitude increases as well and thus successful position control can be re-stabilized. Longer protocols $T_y$ result in an accumulating of numerical truncation errors.

\subsubsection{Orientation control with speed adjustment} \label{subsec:pseudo_orientation}
\begin{figure}
\centering
\includegraphics[width=\linewidth]{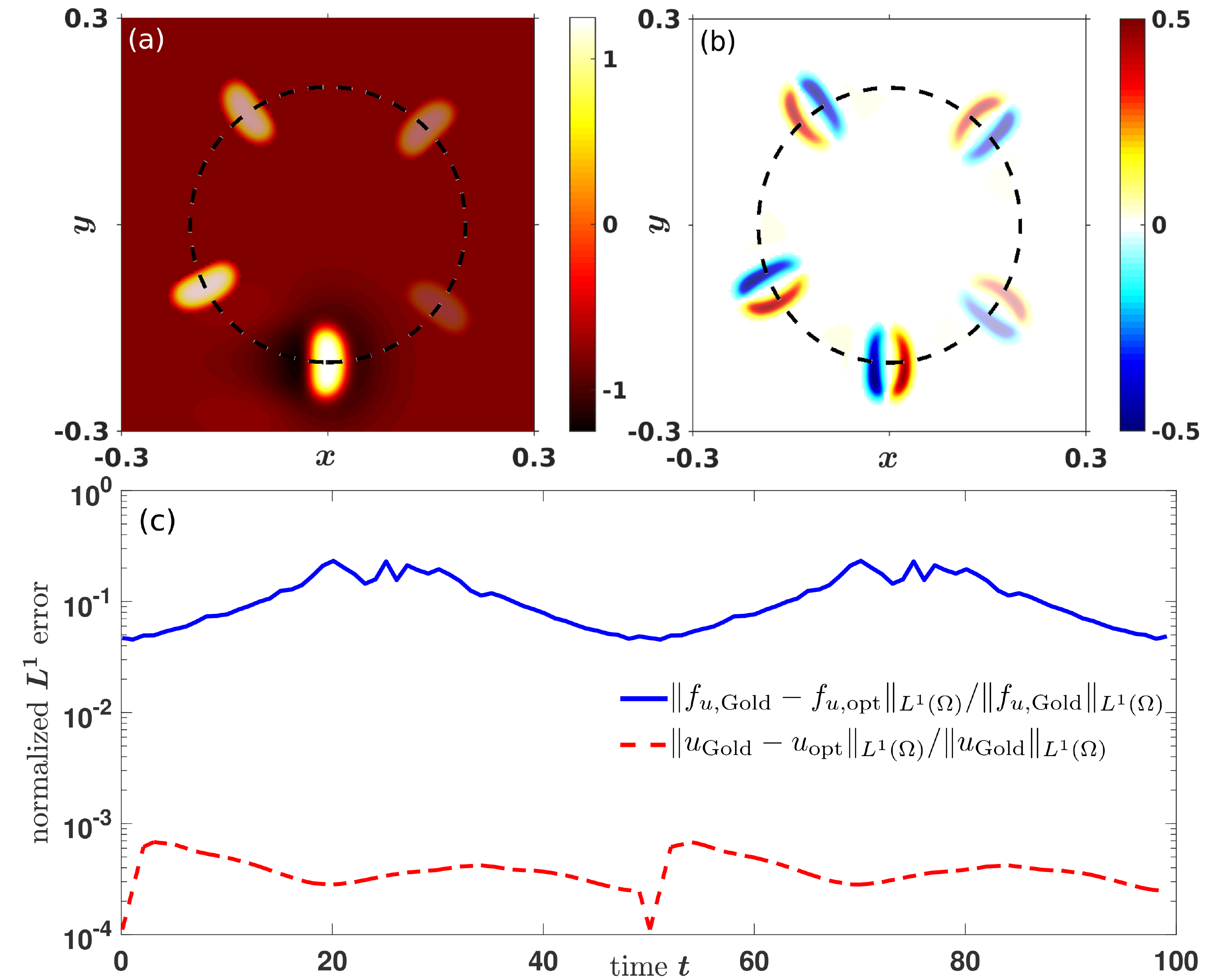}
\caption{Position control along a circular desired distribution \eq{eqs:circle_control} with radius $r=0.2$ and duration time $T=200$; cf.\@ [SI\_video4]. (a) Snapshots of the activator distribution $u(\V{r},t)$ at time moments $t=\{30,75,120,165,200\}$. (b) Control $f_{u,\mathrm{Gold}}(\V{r},t)$ at the same instants of time. In (a) and (b), the dark dashed line indicates the POM and the decreasing transparency marks consecutive moments. (c) Temporal behavior of the relative $L^1(\Omega)$ error \eq{eq:def_l1norm} between expression \eq{eq:fins}, $f_{u,\mathrm{Gold}}$ \eq{eq:fana}, and optimal activator control signals $f_{u,\mathrm{opt}}$ \eq{eq:functopt} during $t \in [0,T/2]$. We select set $1$ in \tabref{tab:simu_parameter} for the kinetic parameters to \eq{eq:spot_model} and Tikhonov parameter is set to $\nu=10^{-7}$. }
\label{fig:position_circle}
\end{figure}

In the previous paragraph, we've demonstrated that the stability of position control can be enhanced if in any current position of the spot its symmetry axis, given by $\Phi_\varphi(t)$, points tangentially to the direction of motion. Therefore, in our next example, we propose to shift the spot pattern along a circular trajectory by simultaneously controlling its orientation
\begin{align}
\Phi_x(t)=\,r\sin\bracket{\Phi_\varphi(t)},\quad
\Phi_y(t)=\,-r\cos\bracket{\Phi_\varphi(t)},\quad
\Phi_\varphi(t) = 2\pi t/T.
\label{eqs:circle_control}
\end{align}
Here, $r$ denotes the radius of the circle and $T$ the protocol's duration. For experimental realization compare \cite{qiao_enhancement_2008}, for example.

In \figref{fig:position_circle}, we present the temporal evolution of the activator distribution $u$ (a) controlled by $\V{f}_\mathrm{Gold}$ (b). In line with the POM, the spot always keeps its symmetry axis at the tangent to the desired trajectory of motion. The control $\V{f}_\mathrm{Gold}$ remains localized and is dominated by the translational Goldstone mode $\partial_{\xi_x} \V{U}_c$ due to the acceleration along $\grb{\Xi}(t)$ \eq{eqs:circle_control}; the average speed is $\bar{v}\simeq 2.4 v_0^x$. Notably, the maximum value of the control magnitude is half as strong compared to position control without adjusting the orientation, $\Phi_\varphi(t)=0$ (not explicitly shown). In panel (c), the temporal behavior of the relative error $\norm{f_{u,\mathrm{Gold}}(t)-f_{u,\mathrm{opt}}(t)}_{\lao}/\norm{f_{u,\mathrm{Gold}}(t)}_{\lao}$ measured by the $L^1\bracket{\Omega}$ norm is shown (solid line). They are large compared to pure position control along a Lissajous curve, cf.\@ \figref{fig:position_lissajous}. Stronger deviations are caused by interpolation errors arising during numerical rotation of spot patterns by $\Phi_\varphi(t)$. The relative error attains a maximum at $\Phi_\varphi(t)=m\,45^\circ,\,m \,\mbox{odd}$. At these angles, the distance between the nodes of the rotated grid and the underlying one is the largest, viz., $dx/\sqrt{2}$, and, hence, numerical interpolation errors become significant. Contrarily, the relative error minimizes at $\Phi_\varphi(t)=m\,90^\circ,\,m \in \Z$, at which both grids coincide. Remarkable, the normalized error $\norm{u_{\mathrm{Gold}}(t)-u_{\mathrm{opt}}(t)}_{\lao}/\norm{u_{\mathrm{Gold}}(t)}_{\lao}$ (dashed line), is still less than $10^{-3}$ at any instants of time despite that the deviation of the associated controls rises up to $\sim 25\%$.

\subsection{Orientation control} \label{sec:orientation_control}
\begin{figure}
\centering
\includegraphics[width=\linewidth]{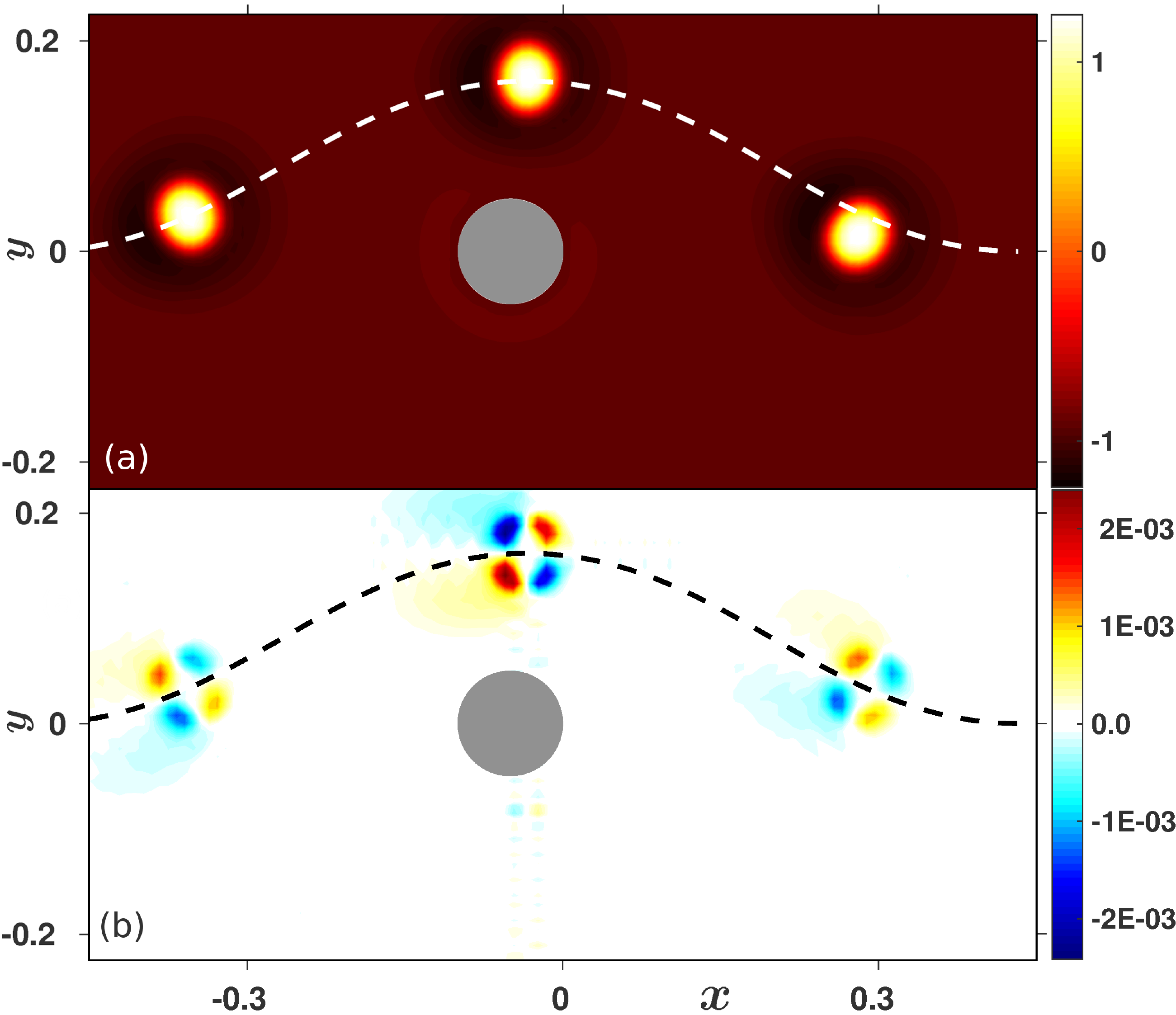}
\caption{Orientation control to avoid collision with circular heterogeneity [SI\_video5]. Snapshots of the activator $u$ (a) controlled by $\V{f}_\mathrm{Gold}$ (b), \eq{eq:fana_orientation}, at different instants of time $t=\{100,250,400\}$. The controlled spot dynamics is simulated on a $(-0.5, 0.5]\times (-0.25, 0.25]$ domain with periodic boundary conditions using ETD2. We use the parameter set 2 in \tabref{tab:simu_parameter}. The circular defect with radius $R=0.05$ is modeled by a jump in $\kappa_1$ from its background value of $\kappa_1^\mathrm{back}=-7.30$ to the value inside the heterogeneity $\kappa_1^\mathrm{het}=-7.50$.}
\label{fig:orientcontrol_heterogenity}
\end{figure}

If the uncontrolled spot propagates at non-zero velocity $\V{v}_0\neq \grb{0}$, the simplest way to navigate it through a spatial domain is to control exclusively its current orientation $\Phi_\varphi(t)$. If so, the translational components of the POM $\grb{\Xi}(t)$ are determined by
\begin{equation} \label{eq:POM_orientation}
\dot{\Phi}_x(t)=\,v_0^x\cos\bracket{\Phi_\varphi(t)},\quad \dot{\Phi}_y(t)=\,v_0^x\sin\bracket{\Phi_\varphi(t)}.
\end{equation}
Clearly, one loses the possibility to control separately the $x$- and $y$-position of the pattern by limiting the speed to $\norm{\dot{\grb{\Phi}}(t)}=v_0^x$. Inserting \eq{eq:POM_orientation} into \eq{eq:fana}, the translational Goldstone modes drop out and we obtain 
\begin{equation} \label{eq:fana_orientation}
\V{f}_\mathrm{Gold}(\V{r},t) =\,-\dot{\Phi}_\varphi(t) \partial_\varphi \V{U}_c(\mathbb{A}(-\Phi_\varphi(t))\bracket{\V{r}-\grb{\Phi}(t)}).
\end{equation}
Now, we pick up the problem formulated in \secref{sec:model}, namely, how to prevent pinning of a spot at a local heterogeneity in the domain. The heterogeneity is viewed as circular region where the parameter $\kappa_1$ jumps from a background value $\kappa_1^\mathrm{back},\,\forall \V{r} \notin \Omega_\circ$ to a defect value $\kappa_1^\mathrm{het},\,\forall \V{r} \in \Omega_\circ$ whereby $\Omega_\circ=\lbrace \bracket{x,y} \in \R^2: (x+R)^2+y^2 < R^2\rbrace$ with radius $R=0.05$. The orientational POM for avoiding the heterogeneity is set to
\begin{equation}
\Phi_\varphi(t)=\frac{\pi}{4}\sin\bracket{\frac{2\pi t}{T}},
\end{equation}
with duration $T=L_x/v_0^x$. Note that the corresponding prescribed positions $\bracket{\Phi_x(t),\Phi_y(t)}^T$ have to be calculated numerically.

In \figref{fig:orientcontrol_heterogenity}, we present the temporal evolution of the activator distribution $u$ in panel (a) and the corresponding control signal $f_{u,\mathrm{Gold}}$ in panel (b). 
The prescribed translational POM is indicated by the dashed lines. At first glance, the control signal possesses a more complicated shape and its magnitude is significantly reduced, 
$ |f_{u,\mathrm{Gold}}|\lesssim 10^{-3}$, as compared to $|f_{u,\mathrm{Gold}}|\lesssim 10^{0}$ and $|f_{u,\mathrm{Gold}}|\lesssim 10^{-1}$ in the previous examples, cf.\@ \figref{fig:position_lissajous} and \figref{fig:position_circle}. 
Thus, orientation control is less invasive than position control. In return, we lose the ability for fast intervention into spot dynamics as well as for mayor increase in the speed of the spot. 
Additionally, orientation control is much more susceptible to fail. The small control magnitudes are too weak to suppress the impact of numerical round-off errors which may result in undesired spot rotation, cf.\@ \secref{subsec:stability}. 
Caused by the small propagation velocity, the duration $T$ of the POM grows as compared to position control, see \secref{subsec:position_control}, and therefore the probability of failure increases as well. 

\subsection{Position control by a single control signal} \label{subsec:singcontrol}

\begin{figure}
\centering
\includegraphics[width=\textwidth]{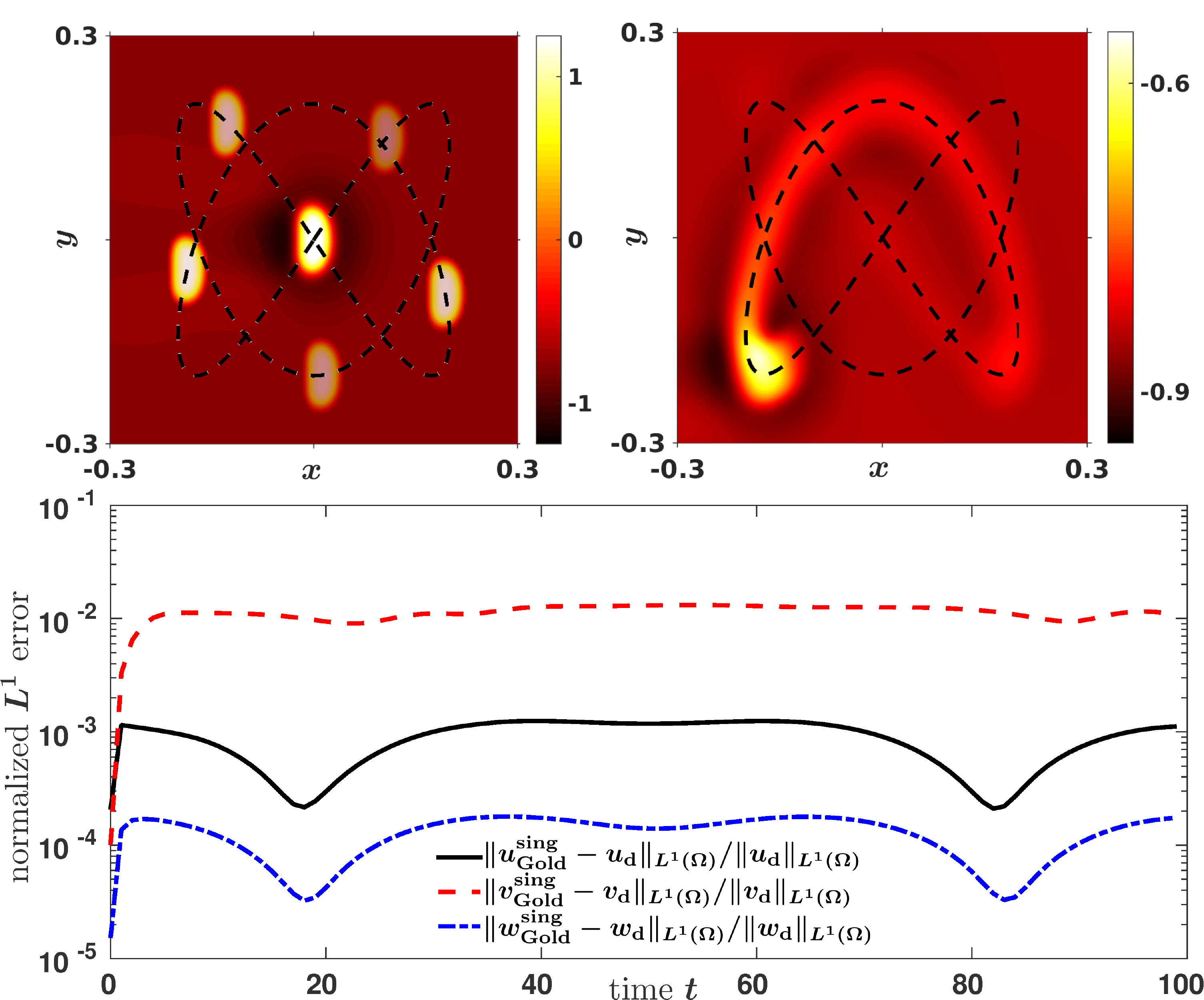}
\caption{Position control by a single control signal acting on $u$, $\V{f}_\mathrm{Gold}^\mathrm{sing}=(f_{u,\mathrm{Gold}}^\mathrm{sing},0,0)^T$, along the Lissajous curve \eq{eq:lissajous_curve}, see [SI\_video6]. (a) Time evolution of activator distribution $u$ at time moments $t=\{10,50,90,130,170,200\}$. The decreasing transparency marks consecutive instants of time. (b) Distribution of the inhibitor $v$ at $t=180$. (c) Temporal behavior of the relative error as measured by the $L^1(\Omega)$ norm \eq{eq:def_l1norm}, between the numerically obtained states $\V{U}$ and the desired distribution $\V{U}_d$ during $t \in [0,T/2]$. We use parameter set $1$ in \tabref{tab:simu_parameter} for the calculations.}
\label{fig:singposition_lissajous}
\end{figure}

So far, we have discussed examples of fully actuated systems for which the number of state components equals the number of independent control signals. If the coupling matrix $\mathbb{B}$ is not invertible, expression \eq{eq:fana} for $\V{f}_\mathrm{Gold}$ cannot be used. The question arises how to extend our approach to underactuated systems \cite{loeber_engel2014,lober2017thesis}. In the following example we assume a control acting on the activator $u$ only while inhibitors $v$ and $w$ remain uncontrolled, i.e., $f_{v,\mathrm{Gold}}(\V{r},t)=f_{w,\mathrm{Gold}}(\V{r},t)=0$. Control via an inhibitor has been discussed in detail for the Hodgkin-Huxley model and the three-component Oregonator model for photosensitive BZ reaction, compare supplemental information to \cite{loeber_engel2014}.\\
To derive an expression for $f_{u,\mathrm{Gold}}(\V{r},t)$, we start with the fully actuated system
\begin{subequations} \label{eq:spot_model2}
\begin{align}
\partial_t u(\V{r},t) =& D_u \Delta u + \kappa_2\,u -u^3- \kappa_3 v -\kappa_4 w + \kappa_1+f_u, \label{eq:spot_model2_u}\\
\tau \partial_t v(\V{r},t) =& D_v \Delta v + u -v+f_v, \label{eq:spot_model2_v}\\
\theta \partial_t w(\V{r},t) =& D_w \Delta w + u -w+f_w. \label{eq:spot_model2_w}
\end{align} 
\end{subequations}
Equations \eq{eq:spot_model2_v}-\eq{eq:spot_model2_w} are linear, inhomogeneous PDEs with initial conditions $v(\V{r},t_0)=v_0(\V{r})$ and $w(\V{r},t_0)=w_0(\V{r})$, respectively. Their solutions can be written as
\begin{align} \label{eq:greenform}
\tilde{v}(\V{r},t) = \mathcal{K}_v^0 \circ v_0+\frac{1}{\tau}\mathcal{K}_v \circ \bracket{u+f_v},\quad  \tilde{w}(\V{r},t) = \mathcal{K}_w^0\circ  w_0+\frac{1}{\theta}\mathcal{K}_w \circ\bracket{u+f_w},
\end{align}
where $\mathcal{K}_i^0$ and $\mathcal{K}_i$, $i \in \lbrace v,w\rbrace$, are integral operators involving Green's functions to the homogeneous equations corresponding to  \eq{eq:spot_model2_v}-\eq{eq:spot_model2_w} with associated initial conditions and to the inhomogeneous equations with zero initial conditions. Plugging \eq{eq:greenform} into \eq{eq:spot_model2_u} gives
\begin{align}\label{eq:spot_model2_u_green}
\partial_t u(\V{r},t) =&\,D_u \Delta u + \kappa_2\,u -u^3- \kappa_3 \left[ \mathcal{K}_v^0 \circ v_0+\frac{1}{\tau}\mathcal{K}_v \circ u \right] -\kappa_4 \left[  \mathcal{K}_w^0 \circ w_0+\frac{1}{\theta}\mathcal{K}_w \circ u \right] + \kappa_1 \notag \\  &\,+f_u - \frac{\kappa_3}{\tau}\mathcal{K}_v \circ f_v - \frac{\kappa_4}{\theta}\mathcal{K}_w \circ f_w.
\end{align}
From the last line of \eq{eq:spot_model2_u_green} we identify the expression for $f_{u,\mathrm{Gold}}(\V{r},t)$ to be
\begin{align}\label{eq:singfana}
f_{u,\mathrm{Gold}}^\mathrm{sing} \bracket{\V{r},t} = f_{u,\mathrm{Gold}} \bracket{\V{r,t}} - \frac{\kappa_3}{\tau}\mathcal{K}_v \circ f_{v,\mathrm{Gold}} - \frac{\kappa_4}{\theta}\mathcal{K}_w \circ f_{w,\mathrm{Gold}},
\end{align}
whereby the component of $\V{f}_\mathrm{Gold}$ are determined by \eq{eq:fana}.

\noindent As an example for position control by a single control signal, we guide a spot along the Lissajous curve given by \eq{eq:lissajous_curve} with radius $r=0.2$ and protocol duration $T=200$. The spot's orientation $\Phi_\varphi(t)=0$ remains uncontrolled. The relative errors between desired and controlled states are shown in \figref{fig:singposition_lissajous}(c). All states are obtained from numerical simulation of \eq{eq:spot_model}-\eq{eq:contRDS} with control $\V{f}_\mathrm{Gold}^\mathrm{sing}=(f_{u,\mathrm{Gold}}^\mathrm{sing},0,0)^T$ given by \eq{eq:singfana}. The relative error for the activator $u$ (solid line) is less than $10^{-3}$ at any time $t$ and thus the controlled activator pattern agrees satisfactorily well with the desired distribution. This finding is corroborated by snapshots of $u$ at different instants of time in \figref{fig:singposition_lissajous}(a). In contrast to the activator, the profile of the inhibitor $v$ is not preserved under control but deformed considerably, see \figref{fig:singposition_lissajous}(b). In particular, an elongated region of activity becomes apparent along the Lissajous curve due to time scale separation in the RDS \eq{eq:spot_model}. The concentration of the slow inhibitor $v$, produced in the wake of the activator, decays exponentially to the rest state on a time scale $\tau=48\approx T/4$. Consequently, the relative error of $v$ (dashed line) attains relatively large values of the order $10^{-2}$. On the other hand, the fast inhibitor $w$ and the activator  $u$ vary on the same characteristic time scale as $\theta=1$ was chosen in the considered example. Thus, we expect only small changes in both profiles in the presence of the control. In fact, the values of the relative error for $w$ turn out to be less than $10^{-4}$ which is even one magnitude smaller than the relative error of $u$, cf.\@ the dash-dotted line in \figref{fig:singposition_lissajous}(c).

\section{Conclusion} \label{sec:conclusion}
% Localized traveling chemical, chemo-mechanical, electrical or neural activity is ubiquitous in spatially extended nonlinear systems driven far from thermodynamic equilibrium. Thus, to control the current position, orientation and velocity of a traveling spot is a key challenge not only under general aspects but particularly from the perspective of numerous applications.

Localized traveling patterns are ubiquitous in spatially extended nonlinear systems driven far from thermodynamic equilibrium. These structures are often coined dissipative solitons or shortly spots and have been observed in various chemical, chemo-mechanical, electrical or neural systems. Hence, to control the position, orientation and velocity of a traveling spot is a key challenge. 
%not only under general aspects but particularly from the perspective of in numerous applications. 

Exploiting the translational and rotational symmetries of the governing equations, we've demonstrated that the control signal, which one has to apply to solve these tasks, is constituted by the Goldstone modes with time dependent prefactors. Intriguingly, for the latter analytic expressions have been derived. To deduce the control signal -- coined Goldstone mode control -- for a given protocol of motion, it is adequate to measure the spot's profile and the corresponding propagation velocity with sufficient accuracy. In particular, Goldstone mode control is realized by external spatio-temporal forcing, i.e., it is an open-loop control. Contrary to closed-loop or feedback control, continuous monitoring of the system is not required. On the downside, as any open-loop control, the method is sensitive to perturbations. Therefore, the range of applicability has been checked by a stability analysis. Most importantly, our approach requires no detailed knowledge about the underlying reaction kinetics as opposed to standard open-loop control.
% as long as the intrinsic propagation velocity and profile of the uncontrolled spot have been measured with sufficient accuracy. 
Although the control signal is invasive, it is designed to preserve simultaneously the shape of the controlled pattern.

% The control acts locally as long as the Goldstone modes are localized and preserves the spatial profile of the uncontrolled spot as long as the spectral gap between deformation and Goldstone modes in the linear stability operator of the spot solution is sufficiently large. Most importantly, to determine the control amplitudes, we do not need any knowledge about the nonlinear kinetics which in many applications are known at best only approximately. In multi-component systems, additionally the coupling matrix must be known.

Remarkably, in all examples considered so far, Goldstone mode control is, within numerical accuracy, equal to solutions of an equivalent, non-regularized optimal control problem. Consequently, our control turns out to be optimal, i.e., no other control enforces the system closer to the desired target state according to the protocol of motion. Furthermore, these control signals have been proven to be excellent initial conditions for regularized optimal control problems; achieving a substantial computational speed-up. Generally, Goldstone mode control approach might serve as consistency check for numerical optimal control algorithms as well. 
We emphasize that optimal control is not only computationally demanding but requires full knowledge of the nonlinear kinetics. On the other hand, the scope of optimal control can be extended in many ways like sparse control or for inequality conditions for the control amplitudes' upper and lower bounds \cite{hlt98,Ryll2016}.
%More general objective functionals than stated above can be studied, compare sparse optimal control \cite{hlt98} acting only in the most sensitive spatial or temporal regions of a controlled solution while vanishing elsewhere. Moreover, optimal control can easily deal with control signals confined to prescribed spatial regions or with upper and lower bounds for the control amplitudes in the form of inequalities, i.e., $-\infty < f_a \leq f_i(\V{r},t) \leq f_b < \infty,\,\forall i=1,\ldots,n$. Technical details as well as examples can be found in \cite{hlt98,Ryll2016}.

Due to the underlying symmetry considerations, Goldstone mode control is widely applicable. Already, the method have been successfully used to guide traveling interfaces and excitation pulses in $1$D \cite{loeber_engel2014,Ryll2016} and spiral waves \cite{Ryll2016} as well as to shape iso-concentration lines of traveling wave patterns \cite{lober2014shaping} in $2$D. Recently, we successfully applied Goldstone mode control to spot solutions of neural field equations \cite{ziepke2018control} that phenomenologically describe the dynamics of synaptically coupled neurons \cite{coombes2014neural}.

\ack
\noindent We thank Alexander Ziepke for helpful discussions as well as for critical reading of the manuscript and acknowledge financial support from the German Science Foundation DFG through the SFB 910 ``Control of Self-Organizing Nonlinear Systems''.

\vspace{3em}
\bibliographystyle{iopart-num_jft2014}
\bibliography{litbank}

\providecommand{\newblock}{}
\begin{thebibliography}{10}
\expandafter\ifx\csname url\endcsname\relax
  \def\url#1{{\tt #1}}\fi
\expandafter\ifx\csname urlprefix\endcsname\relax\def\urlprefix{URL }\fi
\providecommand{\eprint}[2][]{\url{#2}}

\bibitem{kerner2013autosolitons}
Kerner B~S and Osipov V~V 2013 {\em Autosolitons: a new approach to problems of
  self-organization and turbulence\/} vol~61 (Springer Science \& Business
  Media)

\bibitem{purwins_dissipative_2005}
Purwins H~G, B\"{o}deker H~U and Liehr A~W 2005 Dissipative solitons in
  reaction-diffusion systems {\em Dissipative solitons\/} (Springer) pp
  267--308 \urlprefix\url{http://link.springer.com/chapter/10.1007/10928028_11}

\bibitem{Laing2002}
Laing C~R, Troy W~C, Gutkin B and Ermentrout G~B 2002 {\em SIAM J. Appl.
  Math.\/} \href{http://dx.doi.org/10.1137/S0036139901389495}{{\bf 63} 62--97}
  (\textit{Preprint} \eprint{https://doi.org/10.1137/S0036139901389495})
  \urlprefix\url{https://doi.org/10.1137/S0036139901389495}

\bibitem{Purwins2010}
Purwins H~G, B\"odeker H and Amiranashvili S 2010 {\em Adv. Phys.\/}
  \href{http://dx.doi.org/10.1080/00018732.2010.498228}{{\bf 59} 485--701}
  (\textit{Preprint} \eprint{https://doi.org/10.1080/00018732.2010.498228})
  \urlprefix\url{https://doi.org/10.1080/00018732.2010.498228}

\bibitem{arecchi_pattern_1999}
Arecchi F~T, Boccaletti S and Ramazza P 1999 {\em Phys. Rep.\/}
  \href{http://dx.doi.org/10.1016/S0370-1573(99)00007-1}{{\bf 318} 1--83} ISSN
  0370-1573
  \urlprefix\url{http://www.sciencedirect.com/science/article/pii/S0370157399000071}

\bibitem{vanag_localized_2007}
Vanag V~K and Epstein I~R 2007 {\em Chaos\/}
  \href{http://dx.doi.org/10.1063/1.2752494}{{\bf 17} 037110}
  \urlprefix\url{http://scitation.aip.org/content/aip/journal/chaos/17/3/10.1063/1.2752494}

\bibitem{wolff2001spatiotemporal}
Wolff J, Papathanasiou A~G, Kevrekidis I~G, Rotermund H~H and Ertl G 2001 {\em
  Science\/} \href{http://dx.doi.org/10.1126/science.1063597}{{\bf 294}
  134--137} \urlprefix\url{http://science.sciencemag.org/content/294/5540/134}

\bibitem{Sheintuch2008}
Viswanathan G~A, Sheintuch M and Luss D 2008 {\em Ind. Eng. Chem. Res.\/}
  \href{http://dx.doi.org/10.1021/ie8005726}{{\bf 47} 7509--7523}
  (\textit{Preprint} \eprint{https://doi.org/10.1021/ie8005726})
  \urlprefix\url{https://doi.org/10.1021/ie8005726}

\bibitem{le_goff_pattern_2016}
Le~Goff T, Liebchen B and Marenduzzo D 2016 {\em Phys. Rev. Lett.\/}
  \href{http://dx.doi.org/10.1103/PhysRevLett.117.238002}{{\bf 117} 238002}
  \urlprefix\url{http://link.aps.org/doi/10.1103/PhysRevLett.117.238002}

\bibitem{taube_persistent_2003}
Taube J~S and Bassett J~P 2003 {\em Cereb. Cortex\/}
  \href{http://dx.doi.org/10.1093/cercor/bhg102}{{\bf 13} 1162--1172} ISSN
  1047-3211, 1460-2199
  \urlprefix\url{http://cercor.oxfordjournals.org/content/13/11/1162}

\bibitem{gilad_ecosystem_2004}
Gilad E, von Hardenberg J, Provenzale A, Shachak M and Meron E 2004 {\em Phys.
  Rev. Lett.\/} \href{http://dx.doi.org/10.1103/PhysRevLett.93.098105}{{\bf
  93}} \urlprefix\url{http://link.aps.org/doi/10.1103/PhysRevLett.93.098105}

\bibitem{Mikhailov2006}
Mikhailov A and Showalter K 2006 {\em Phys. Rep.\/}
  \href{http://dx.doi.org/https://doi.org/10.1016/j.physrep.2005.11.003}{{\bf
  425} 79--194} ISSN 0370-1573
  \urlprefix\url{http://www.sciencedirect.com/science/article/pii/S0370157305004825}

\bibitem{vanag_design_2008}
Vanag V~K and Epstein I~R 2008 {\em Chaos\/}
  \href{http://dx.doi.org/10.1063/1.2900555}{{\bf 18} 026107}
  \urlprefix\url{http://scitation.aip.org/content/aip/journal/chaos/18/2/10.1063/1.2900555}

\bibitem{Zykov2003}
Zykov V~S, Bordiougov G, Brandtst\"adter H, Gerdes I and Engel H 2003 {\em
  Phys. Rev. E\/} \href{http://dx.doi.org/10.1103/PhysRevE.68.016214}{{\bf 68}
  016214} \urlprefix\url{http://link.aps.org/doi/10.1103/PhysRevE.68.016214}

\bibitem{Chen2009}
Chen J~X, Zhang H and Li Y~Q 2009 {\em J. Chem. Phys.\/}
  \href{http://dx.doi.org/10.1063/1.3098543}{{\bf 130} 124510}
  \urlprefix\url{http://scitation.aip.org/content/aip/journal/jcp/130/12/10.1063/1.3098543}

\bibitem{Haas1995}
Haas G, B\"ar M, Kevrekidis I~G, Rasmussen P~B, Rotermund H~H and Ertl G 1995
  {\em Phys. Rev. Lett.\/}
  \href{http://dx.doi.org/10.1103/PhysRevLett.75.3560}{{\bf 75} 3560}
  \urlprefix\url{https://link.aps.org/doi/10.1103/PhysRevLett.75.3560}

\bibitem{ziepke_wave_2016}
Ziepke A, Martens S and Engel H 2016 {\em J. Chem. Phys.\/}
  \href{http://dx.doi.org/10.1063/1.4962173}{{\bf 145} 094108} ISSN 0021-9606
  \urlprefix\url{http://aip.scitation.org/doi/abs/10.1063/1.4962173}

\bibitem{pierre_controlling_1996}
Pierre T, Bonhomme G and Atipo A 1996 {\em Phys. Rev. Lett.\/}
  \href{http://dx.doi.org/10.1103/PhysRevLett.76.2290}{{\bf 76} 2290--2293}
  \urlprefix\url{http://link.aps.org/doi/10.1103/PhysRevLett.76.2290}

\bibitem{luthje_control_2001}
L\"{u}thje O, Wolff S and Pfister G 2001 {\em Phys. Rev. Lett.\/}
  \href{http://dx.doi.org/10.1103/PhysRevLett.86.1745}{{\bf 86} 1745--1748}
  \urlprefix\url{http://link.aps.org/doi/10.1103/PhysRevLett.86.1745}

\bibitem{Kim2001}
Kim M, Bertram M, Pollmann M, von Oertzen A, Mikhailov A~S, Rotermund H~H and
  Ertl G 2001 {\em Science\/}
  \href{http://dx.doi.org/10.1126/science.1059478}{{\bf 292} 1357} ISSN
  0036-8075 \urlprefix\url{http://science.sciencemag.org/content/292/5520/1357}

\bibitem{gurevich_instabilities_2013}
Gurevich S~V and Friedrich R 2013 {\em Phys. Rev. Lett.\/}
  \href{http://dx.doi.org/10.1103/PhysRevLett.110.014101}{{\bf 110}}
  \urlprefix\url{http://link.aps.org/doi/10.1103/PhysRevLett.110.014101}

\bibitem{bryson1975applied}
Bryson A~E 1975 {\em Applied optimal control: optimization, estimation and
  control\/} (CRC Press)

\bibitem{hllst02}
Hoffmann K~H, Lasiecka I, Leugering G, Sprekels J and Tr\"oltzsch F (eds) 2002
  {\em Optimal Control of Complex Structures\/} ({\em ISNM\/} vol 139)
  (Birkh\"auser Verlag)

\bibitem{hlt98}
Hoffmann K~H, Leugering G and Tr\"oltzsch F (eds) 1998 {\em Optimal Control of
  Partial Differential Equations\/} ({\em ISNM\/} vol 133) (Birkh\"auser
  Verlag)

\bibitem{bjt10}
Barthel W, John C and Tr\"oltzsch F 2010 {\em Z. Angew. Math. und Mech.\/}
  \href{http://dx.doi.org/10.1002/zamm.200900359}{{\bf 90} 966--982}

\bibitem{casas_ryll_troeltzsch2014}
Casas E, Ryll C and Tr{\"o}ltzsch F 2013 {\em Comp. Meth. Appl. Math.\/}
  \href{http://dx.doi.org/10.1515/cmam-2013-0016}{{\bf 13} 415--442}

\bibitem{WolffPRL2003}
Wolff J, Papathanasiou A~G, Rotermund H~H, Ertl G, Li X and Kevrekidis I~G 2003
  {\em Phys. Rev. Lett.\/}
  \href{http://dx.doi.org/10.1103/PhysRevLett.90.018302}{{\bf 90} 018302}
  \urlprefix\url{http://link.aps.org/doi/10.1103/PhysRevLett.90.018302}

\bibitem{qiao_enhancement_2008}
Qiao L, Li X, Kevrekidis I~G, Punckt C and Rotermund H~H 2008 {\em Phys. Rev.
  E\/} \href{http://dx.doi.org/10.1103/PhysRevE.77.036214}{{\bf 77} 036214}
  \urlprefix\url{http://link.aps.org/doi/10.1103/PhysRevE.77.036214}

\bibitem{steinbock1993control}
Steinbock O, Zykov V~S and M{\"u}ller S~C 1993 {\em Nature\/}
  \href{http://dx.doi.org/10.1038/366322a0}{{\bf 366} 322--324}

\bibitem{Schrader1995}
Schrader A, Braune M and Engel H 1995 {\em Phys. Rev. E\/}
  \href{http://dx.doi.org/10.1103/PhysRevE.52.98}{{\bf 52} 98}
  \urlprefix\url{https://link.aps.org/doi/10.1103/PhysRevE.52.98}

\bibitem{odent_optical_2016}
Odent V, Louvergneaux E, Clerc M~G and Andrade-Silva I 2016 {\em Phys. Rev.
  E\/} \href{http://dx.doi.org/10.1103/PhysRevE.94.052220}{{\bf 94} 052220}
  \urlprefix\url{http://link.aps.org/doi/10.1103/PhysRevE.94.052220}

\bibitem{Zykov2004}
Zykov V~S, Bordiougov G, Brandtst\"adter H, Gerdes I and Engel H 2004 {\em
  Phys. Rev. Lett.\/}
  \href{http://dx.doi.org/10.1103/PhysRevLett.92.018304}{{\bf 92} 018304}
  \urlprefix\url{http://link.aps.org/doi/10.1103/PhysRevLett.92.018304}

\bibitem{Schlesner2008book}
Schlesner J, Zykov V and Engel H 2008 {\em Handbook of Chaos Control\/}
  (Wiley-VCH Verlag) chap Feedback-mediated Control of Hypermeandering Spiral
  Waves, pp 591--607 ISBN 9783527622313
  \urlprefix\url{http://dx.doi.org/10.1002/9783527622313.ch27}

\bibitem{sakurai2002}
Sakurai T, Mihaliuk E, Chirila F and Showalter K 2002 {\em Science\/}
  \href{http://dx.doi.org/10.1126/science.1071265}{{\bf 296} 2009--2012} ISSN
  0036-8075 \urlprefix\url{http://science.sciencemag.org/content/296/5575/2009}

\bibitem{STotz2018}
Totz S, L\"ober J, Totz J~F and Engel H 2018 {\em NJP\/} {\bf 20} 053034
  \urlprefix\url{http://stacks.iop.org/1367-2630/20/i=5/a=053034}

\bibitem{loeber_engel2014}
L{\"o}ber J and Engel H 2014 {\em Phys. Rev. Lett.\/}
  \href{http://dx.doi.org/10.1103/PhysRevLett.112.148305}{{\bf 112} 148305}
  \urlprefix\url{https://link.aps.org/doi/10.1103/PhysRevLett.112.148305}

\bibitem{buch_eng_kamm_tro2013}
Buchholz R, Engel H, Kammann E and Tr\"oltzsch F 2013 {\em Comput. Optim.
  Appl.\/} \href{http://dx.doi.org/10.1007/s10589-013-9550-y}{{\bf 56}
  153--185}
  \urlprefix\url{https://link.springer.com/article/10.1007/s10589-013-9550-y}

\bibitem{LoeberBook2014}
L\"ober J, Coles R, Siebert J, Engel H and Sch\"oll E 2015 Control of chemical
  wave propagation {\em Engineering of Chemical Complexity II\/} ed Mikhailov A
  and Ertl G (Singapore: World Scientific)

\bibitem{lober2014stability}
L{\"o}ber J 2014 {\em Phys. Rev. E\/}
  \href{http://dx.doi.org/10.1103/PhysRevE.89.062904}{{\bf 89} 062904}
  \urlprefix\url{https://link.aps.org/doi/10.1103/PhysRevE.89.062904}

\bibitem{lober2014shaping}
L{\"o}ber J, Martens S and Engel H 2014 {\em Phys. Rev. E\/}
  \href{http://dx.doi.org/10.1103/PhysRevE.90.062911}{{\bf 90} 062911}
  \urlprefix\url{https://link.aps.org/doi/10.1103/PhysRevE.90.062911}

\bibitem{Ryll2016}
Ryll C, L{\"o}ber J, Martens S, Engel H and Tr{\"o}ltzsch F 2016 Analytical,
  optimal, and sparse optimal control of traveling wave solutions to
  reaction-diffusion systems {\em Control of Self-Organizing Nonlinear
  Systems\/} ed Sch{\"o}ll E, Klapp S~H~L and H{\"o}vel P (Springer) pp
  189--210 ISBN 978-3-319-28028-8
  \urlprefix\url{http://dx.doi.org/10.1007/978-3-319-28028-8_10}

\bibitem{doelman_pulse_2009}
Doelman A, van Heijster P and Kaper T~J 2009 {\em J. Dyn. Differ. Equ.\/}
  \href{http://dx.doi.org/10.1007/s10884-008-9125-2}{{\bf 21} 73--115}
  \urlprefix\url{http://link.springer.com/10.1007/s10884-008-9125-2}

\bibitem{van_heijster_pulse_2008}
van Heijster P, Doelman A and Kaper T~J 2008 {\em Physica D\/}
  \href{http://dx.doi.org/10.1016/j.physd.2008.07.014}{{\bf 237} 3335--3368}
  \urlprefix\url{http://linkinghub.elsevier.com/retrieve/pii/S0167278908002923}

\bibitem{nishiura2011}
Nishiura Y, Teramoto T and Yuan X 2011 {\em Comm. Pure Appl. Anal.\/}
  \href{http://dx.doi.org/10.3934/cpaa.2012.11.307}{{\bf 11} 307--338} ISSN
  1534-0392
  \urlprefix\url{http://www.aimsciences.org/journals/displayArticlesnew.jsp?paperID=6492}

\bibitem{gurevich_breathing_2006}
Gurevich S~V, Amiranashvili S and Purwins H~G 2006 {\em Phys. Rev. E\/}
  \href{http://dx.doi.org/10.1103/PhysRevE.74.066201}{{\bf 74} 066201}
  \urlprefix\url{http://link.aps.org/doi/10.1103/PhysRevE.74.066201}

\bibitem{yang_jumping_2006}
Yang L, Zhabotinsky A~M and Epstein I~R 2006 {\em Phys. Chem. Chem. Phys.\/}
  \href{http://dx.doi.org/10.1039/B609214D}{{\bf 8} 4647--4651}
  \urlprefix\url{http://xlink.rsc.org/?DOI=B609214D}

\bibitem{marwaha_hot_2003}
Marwaha B and Luss D 2003 {\em Chem. Eng. Sci\/}
  \href{http://dx.doi.org/10.1016/S0009-2509(02)00602-4}{{\bf 58} 733--738}
  \urlprefix\url{http://linkinghub.elsevier.com/retrieve/pii/S0009250902006024}

\bibitem{mikhailov1983stochastic}
Mikhailov A, Schimansky-Geier L and Ebeling W 1983 {\em Phys. Lett. A\/}
  \href{http://dx.doi.org/10.1016/0375-9601(83)90163-9}{{\bf 96} 453--456}
  \urlprefix\url{http://www.sciencedirect.com/science/article/pii/0375960183901639}

\bibitem{Nishiura2005}
Nishiura Y, Teramoto T and Ueda K~I 2005 {\em Chaos\/}
  \href{http://dx.doi.org/10.1063/1.2087127}{{\bf 15} 047509}
  \urlprefix\url{https://aip.scitation.org/doi/abs/10.1063/1.208712}

\bibitem{Biktashev2015}
Biktasheva I~V, Dierckx H and Biktashev V~N 2015 {\em Phys. Rev. Lett.\/}
  \href{http://dx.doi.org/10.1103/PhysRevLett.114.068302}{{\bf 114}(6) 068302}
  \urlprefix\url{http://link.aps.org/doi/10.1103/PhysRevLett.114.068302}

\bibitem{Casas2018}
Casas E, Mateos M and R\"osch A 2018 {\em Comp. Opt. Appl.\/}
  \href{http://dx.doi.org/10.1007/s10589-018-9979-0}{{\bf 70} 239--266}
  \urlprefix\url{https://doi.org/10.1007/s10589-018-9979-0}

\bibitem{lober2017thesis}
L{\"o}ber J 2017 {\em Control of Reaction-Diffusion Systems\/} (Cham: Springer
  International Publishing) pp 195--220 ISBN 978-3-319-46574-6
  \urlprefix\url{http://dx.doi.org/10.1007/978-3-319-46574-6_5}

\bibitem{ziepke2018control}
Ziepke A, Martens S and Engel H 2018 {\em arXiv:1806.10938\/}

\bibitem{coombes2014neural}
Coombes S, beim Graben P, Potthast R and Wright J 2014 {\em Neural Fields\/}
  (Springer-Verlag Berlin Heidelberg) ISBN 978-3-642-54593-1

\end{thebibliography}

\end{document}